\documentclass{article}

\usepackage{amsfonts}
\usepackage{amsthm,amsmath,amssymb,accents,empheq} 
\usepackage{graphicx} % Required for inserting images
\usepackage[margin=0.5in]{geometry}
\usepackage{xcolor}
\usepackage[
backend=biber,
style=numeric,
sorting=none
]{biblatex}
\usepackage{csquotes}
\usepackage{appendix}
\usepackage{hyperref}
\usepackage{algorithm}
\usepackage{cancel}
\usepackage[noend]{algpseudocode}
\usepackage{booktabs}

\usepackage{tikz}
\usetikzlibrary{calc}
\usetikzlibrary{decorations.pathreplacing}

\definecolor{DarkPurple}{RGB}{79, 40, 117}
\definecolor{DarkGreen} {RGB}{ 0, 71,  0}
\definecolor{DarkBlue}  {RGB}{ 0,  0, 201}
\definecolor{DarkRed}   {RGB}{ 205,  0, 0}

\title{An Implicit, Moment-Conserving Low-Rank Tensor Method for the Hybrid Kinetic-Ion Fluid-Electron Plasma Model}
\author{Stephen R. White \thanks{Corresponding author: \href{mailto:swhite@lanl.gov}{swhite@lanl.gov}.}\textsuperscript{$\,\,\,$,}\thanks{Los Alamos National Laboratory, Los Alamos, NM, 87544, United States.}
\and Adam J. Stanier \footnotemark[2]
\and Luis Chac\'{o}n \thanks{Los Alamos National Laboratory (Retired), Los Alamos, NM, 87544, United States.}
\and Will T. Taitano \footnotemark[2]
\and Isaac A. Castro \thanks{University of Delaware, Newark, DE, 19716, United States.}
}

\begin{document}

\maketitle

\begin{abstract}
    Quasi-neutral hybrid models for simulating plasmas, with kinetic ions and fluid electrons, have shown great promise for approaching the multi-scale nature of plasmas. In this paper we present a novel hybrid scheme for electromagnetic settings that conserves mass, momentum, and energy, employing a low-rank tensor method utilizing the Tensor Train decomposition to avoid the \enquote{curse of dimensionality.} Electrons are simulated as a continuous fluid, while the kinetic ions are modeled as a distribution function governed by the Vlasov equation. The Vlasov equation is updated with the implicit midpoint method. We use a modified version of a conservative truncation method to guarantee conservation of mass, momentum, and energy throughout the method, which corrects the truncated distribution function without needing to advance a separate set of fluid-moment equations. A series of electrostatic and electromagnetic test problems is used to demonstrate the correctness of the method and its conservation properties.
\end{abstract}
\section{Introduction} \label{section:introduction}

Multi-scale plasmas may be simulated by hybrid plasma models, in which the electrons, being substantially lighter than their companion ions and therefore evolving on a shorter timescale, are described by a fluid model, while the slower ions are modeled as a kinetic species by the Vlasov equation. The hybrid kinetic-ion fluid-electron plasma model is typically run with ion-scale resolution, making it more computationally tractable than a purely kinetic solve, and has seen use in a variety of contexts \cite{Winske_etal_2003, Winske_etal_2023, palmroth_etal_2025, vonAlfthan_2014, stanier_2020}.

There are a number of approaches to solving the Vlasov equation \cite{palmroth_etal_2025, filbet_sonnendrucker_2003_b}. The fully Lagrangian Particle-in-Cell (PIC) method eschews simulating the Vlasov equation as a mesh and opts to simulate a Lagrangian representation of the distribution function instead, that is, as a set of macroparticles evolved along the characteristic lines of the Vlasov equation \cite{Hockney_Eastwood_1988, Birdsall_Langdon_1991, Verboncoeur_2005}. This method is effective and avoids representing the distribution function as a grid; however, it is known to suffer from discrete particle noise that scales as $\mathcal{O}\left(N_p^{-1/2}\right)$, where $N_p$ is the number of macroparticles, as well as aliasing errors from the mismatch between particles in continuous space and fields in discrete space, which manifest as dispersion issues \cite{stanier_2020} or finite-grid instabilities, and can induce numerical heating or cooling \cite{Birdsall_Langdon_1991}. Despite these drawbacks, it remains one of the most successful and flexible approaches to plasma simulation. Energy-conserving PIC schemes have been studied thoroughly in the last decade, with implicit \cite{Chen_etal_2011, Chen_chacon_2014, Chen_Chacon_2025}, semi-implicit \cite{Markidis_2011, Lapenta_2017, Chen_Chacon_etal_2020}, and even more recently explicit \cite{Ricketson_2025} updates. PIC has been used as the solver for the kinetic component of the hybrid model \cite{stanier2019fully, stanier_2020, stanier_chacon_2022, ari_etal_2023}. 

As an alternative to PIC, we may impose a grid structure upon the distribution function. To do so, we have two main lines of approach, Semi-Lagrangian (SL) and Eulerian. SL methods represent the distribution function as a grid and evolve the grid points along characteristics in a manner analogous to PIC \cite{cheng_1976, sonnendrucker_1999, Dimarco_2014, Crouseilles_2009}. High-order conservative SL methods for Vlasov-Poisson using WENO were developed in the 2010s and 2020s \cite{Qiu_Christlieb_2010, Cai_2016, Zheng_2022}, with Ref.\ \cite{Xu_2026} achieving conservative SL for Vlasov-Maxwell by cascade remapping. Recent work has developed mass-, momentum-, and energy-conserving SL methods for Vlasov-Amp\`{e}re \cite{Liu_2023} and Vlasov-Maxwell \cite{Liu_2025}. More notably for us, the recent preprint \cite{zheng_sands_qiu_2026} achieved a low-rank SL method for Vlasov-Poisson that conserves mass, momentum, and energy as well. Ref.\ \cite{Einkemmer_2017} gives a good overview of structure-preserving SL methods up through 2017. SL methods have seen use in the hybrid context \cite{Dolgov_2014, Rojas_2021, Wang_2022, Zhou_etal_2022, Li_etal_2026}.\footnote{Notably, Ref.\ \cite{Dolgov_2014} employs a low-rank implementation of Tensor Trains, with Ref.\ \cite{Rojas_2021} mentioning ambitions to do the same.} Among these, Ref.\ \cite{Li_etal_2026} does so while conserving mass, momentum, and energy.

Eulerian methods for solving the Vlasov equation \cite{Shoucri_2008, Arber_Vann_2002, Filbet_Sonnendrucker_2003_a, Crouseilles_etal_2008_pt_1, Crouseilles_etal_2008_pt_2} offer an alternative to tracing characteristics. Refs.\ \cite{Cheng_2014_VA, Cheng_2014_VM} developed both explicit and implicit energy-conserving Eulerian solvers for the non-relativistic Vlasov-Amp\`{e}re and Vlasov-Maxwell system, respectively. Eulerian solvers have been used in the hybrid model as well \cite{Valentini_2007, Servidio_2012, vonAlfthan_2014, Crestetto_etal_2022, Taitano_2018, Taitano_2021, Taitano_2021_eulerian_vlasov-fokker-planck-spherical-implosion}. 

While implicit field solves for hybrid models have been studied for some time \cite{Hewett_1980, Lipatov_2002}, there has been exploration into fully implicit methods for both the field and kinetic solvers. For example, Refs.\ \cite{Chen_etal_2011, Cheng_2013, Chen_chacon_2014, Sturdevant_2016, Sturdevant_2017, stanier2019fully, stanier_chacon_2022} all use an implicit particle advance in the context of PIC, with Refs.\ \cite{Cheng_2013, Sturdevant_2016, Sturdevant_2017, stanier2019fully, stanier_chacon_2022} being in the context of hybrid methods. Though rarer due to the computational complexity, implicit Eulerian solves of the Vlasov equation have also been explored, including the Vlasov-Amp\`{e}re system \cite{Anderson_2020, Cheng_2014_VA}, Vlasov-Amp\`{e}re-Fokker-Planck \cite{Taitano_2015_pt_1, Taitano_2015_pt_2}, Vlasov-Poisson \cite{Camporeale_2016}, and Vlasov-Maxwell \cite{Delzanno_2015, Cheng_2014_VM, Shiroto_2019}. Ref.\ \cite{Taitano_2016} developed a conservative and implicit $0$D$2$V scheme for the Vlasov-Fokker-Planck system under a hybrid setting, extending this to $1$D$2$V \cite{Taitano_2018}, adding a thermal-speed-based adaptive grid strategy \cite{Taitano_2021}, and extending this to spherical geometry to apply it to inertial confinement fusion \cite{Taitano_2021_eulerian_vlasov-fokker-planck-spherical-implosion}.

At the core of any sort of grid method is the need to store high-dimensional data ($6$D plus time), and here one encounters the \enquote{Curse of Dimensionality} (CoD), the exponential increase in memory and runtime with the number of dimensions in a system. Storing the distribution function evolved by the Vlasov equation has cost $\mathcal{O}\left(N^{d_x + d_v}\right)$, where $d_x$ and $d_v$ are the dimensions of space and velocity, respectively, and $N$ is the number of elements in each dimension, here assumed uniform for simplicity. For a full $3$D$3$V system with large $N$, the storage alone becomes intractable: with $100$ nodes in every dimension, storing one double ($8$ bytes) at each node, one mesh at one timestep requires roughly $8\times10^{12}$ bytes, or $8$ terabytes, to say nothing of auxiliary variables such as the fields.

The tensor object,\footnote{We note that for the purposes of this paper, a tensor is simply a generalized matrix, a $d$-dimensional object containing an element (here a real number) at each index, carrying with it no special transformation properties.} equipped with a low-rank representation, is a natural fit. Though developed over a century ago, it has recently seen a significant increase in use across a variety of fields \cite{Kolda_Bader_2009}, largely due to its ability to not only store but effectively compress large quantities of data. Of the many tensor decompositions, those most used in the context of Vlasov solvers are the Hierarchical Tucker (HT) decomposition \cite{Hackbusch_Kuhn_2009, Grasedyck_2010} and the Tensor Train (TT) decomposition \cite{oseledets_2011}. 

The Dynamical Low-Rank Approximation, developed in 2007 \cite{Koch_Lubich_2007}, starts with a low-rank representation of the tensor and confines the system dynamics to a low-rank manifold, fixing the rank (though this may be made adaptive). With its subsequent time integrator developed in 2014 \cite{Lubich_Oseledets_2014}, it quickly saw much use in Vlasov solvers, with Refs.\ \cite{Einkemmer_Lubich_2018, Einkemmer_Lubich_2019} applying it to Vlasov-Poisson, Ref.\ \cite{Einkemmer_Ostermann_Piazzola_2020} extending it from Vlasov-Poisson to Vlasov-Maxwell, and Ref.\ \cite{Einkemmer_Joseph_2021} making it conserving.

An alternative to this is the step-and-truncate (SAT) approach, which allows the dynamics of the system to increase the rank, which is then truncated to keep memory in check. Refs.\ \cite{guo_qiu_2022, guo_qiu_2024_1, guo_qiu_2024_2} explored using the HT decomposition as a solution to the CoD for Vlasov-Poisson while also conserving mass, momentum, and energy using a weighted truncation scheme. Ref.\ \cite{zheng_sands_qiu_2026} extended this work by updating the weight functions for its conserving truncation scheme, and another recent preprint explored it in the context of Vlasov-Maxwell \cite{heenatigala_guo_2026}. Ref.\ \cite{Zheng_etal_2025} examined this method in an SL setting for Vlasov-Poisson. An initial investigation into using the TT decomposition method applied to the Vlasov equation, also in the electrostatic context, may be found in Ref.\ \cite{Kormann_2015}. Ref.\ \cite{AllmannRahn_2022} uses the HT to simulate a full $3$D$3$V Vlasov-Maxwell system while preserving mass. Ref.\ \cite{Wang_Hu_2026} explored the use of the TT format in a number of settings, including the Vlasov-Amp\`{e}re-Fokker-Planck equation. Ref.\ \cite{Chikitkin_2021} applied the TT format to the Boltzmann equation for gas dynamics, utilizing a hybrid tensor method in which a $3$V tensor object was stored at each physical location, showcasing an ability to handle nontrivial geometry. Another preprint \cite{chinomona_etal_2026} considers the Vlasov-Maxwell system using TTs to represent Discontinuous Galerkin operators. A recent preprint \cite{nakao_2026} developed a conservative low-rank hybrid method for the electrostatic Vlasov-Fokker-Planck with an implicit-explicit integrator. 

Many of the phenomena for which the hybrid model is suited, such as the linear growth of kinetic instabilities from small seed perturbations, Landau damping, and ion-scale turbulence, involve low-amplitude fluctuations or subtle velocity-space structure, where the discrete particle noise of PIC is most damaging; this has motivated the exploration of Eulerian hybrid-Vlasov solvers. Eulerian solvers trade this noise for the CoD, for which low-rank tensor formats are particularly well suited. A standard SAT truncation step, however, does not preserve the moments of the distribution function, which in a hybrid model has implications beyond the ions alone: the ion kinetic energy is exchanged with the electron pressure and the magnetic field, so energy spuriously injected or removed by truncation can find its way into the fluid and field quantities. The magnetized dynamics of interest, such as ion cyclotron, whistler, and Alfv\'enic waves relevant to magnetic confinement fusion (MCF) \cite{stanier_chacon_2022}, exist only in the electromagnetic setting, necessitating a fully electromagnetic treatment. Under the hybrid model, resolving dispersive whistler waves becomes a challenge: the timestep $\Delta t$ of any explicit update must scale with $\Delta x^2$ \cite{Kunz_Lesur_2013}, where $\Delta x$ is the grid spacing, and explicit hybrid codes commonly resort to subcycling the field update to cope \cite{Matthews_1994, Karimabadi_etal_2004}. An implicit time integrator removes this restriction and permits timesteps set by the ion dynamics of interest, while the low-rank representation mitigates the per-iteration cost that has historically posed a challenge to implicit Eulerian Vlasov solves. 

To date no existing low-rank Vlasov solver combines these ingredients: no fully implicit Eulerian solve has used the HT or TT decomposition in an SAT framework; no conservative TT method has been extended to the full electromagnetic setting; and no HT or TT Eulerian Vlasov solver has been applied to the hybrid model. This paper addresses these lacunae with a fully implicit, conservative, low-rank TT solver for the electromagnetic hybrid model, and proceeds as follows. Section \ref{section:hybrid-model} will set up the system we wish to simulate and discuss its conservative properties in the continuous setting. This will lead to a discretization of the model, where we will show those same properties carry through. Following this, we will engage with a high-level discussion of Tensor Trains in Section \ref{section:tensor-trains} and how we may truncate them in a way that conserves mass, momentum, and energy without solving a separate set of fluid-moment equations, building off the work of Refs.\ \cite{guo_qiu_2022,guo_qiu_2024_1,guo_qiu_2024_2,heenatigala_guo_2026} while reducing rank contribution. All of this will inform Section \ref{section:implementation}, where we will tie all these components together, demonstrating the resulting method by a number of numerical experiments in Section \ref{section:numerical-results}. Finally, we will wrap up with some concluding remarks in Section \ref{section:conclusion}.
\section{The Hybrid Kinetic-Ion Fluid-Electron Model} \label{section:hybrid-model}

\subsection{The Continuous Hybrid Model} \label{subsec:hybrid-vlasov-maxwell}

We are interested in simulating our plasma using a hybrid method in which the ions are viewed kinetically and the electrons are viewed as a bulk fluid.\footnote{See Chapter 5 of Ref.\ \cite{Bittencourt} for the derivation of the Vlasov equation and Chapter 9 of same for an excellent discussion on deriving the magnetohydrodynamic equations and the generalized Ohm's law.} Following in the footsteps of \cite{stanier2019fully}, we choose a simple massless electron fluid with scalar electron pressure. Often the electron pressure is derived from an adiabatic equation of state, $p_e = T_{e0}n^\gamma$, the special case of $\gamma=1$ giving the isothermal scenario \cite{vonAlfthan_2014}. However, a number of applications of our method such as MCF \cite{stanier_chacon_2022} and ICF \cite{Taitano_2021} require resistive and viscous terms in Ohm's law, as well as electron heat conduction. Accordingly, the electron pressure must be evolved to include the electron heat flux as well as the Ohmic and viscous heating; otherwise the energy dissipated by resistivity and viscosity would be lost and total energy would not be conserved. The terms of the system may be referenced in Table \ref{tab:model-symbol-lookup}.

\begin{align}
    \frac{\partial f_s}{\partial t} + \mathbf{v}\cdot\nabla_xf_s + \frac{q_s}{m_s}\left(\mathbf{E}^* + \mathbf{v}\times\mathbf{B}\right)\cdot\nabla_vf_s = 0, \label{eq:vlasov-s} \\
    \frac{\partial \mathbf{B}}{\partial t} = -\nabla\times\mathbf{E}, \label{eq:curl-E} \\
    \mathbf{E} = \mathbf{E}^* + \eta \mathbf{J} \equiv -\mathbf{u}_i\times\mathbf{B} + \frac{\mathbf{J}\times\mathbf{B}}{ne} - \frac{\nabla p_e}{ne} - \frac{\nabla\cdot\overset\leftrightarrow{\Pi}_e}{ne} + \eta\mathbf{J}, \label{eq:E-definition} \\
    \left(\gamma - 1\right)^{-1}\left[\frac{\partial p_e}{\partial t} + \nabla\cdot\left(\mathbf{u}_ep_e\right)\right] + p_e\nabla\cdot\mathbf{u}_e = H_e - \nabla\cdot\mathbf{q}_e, \label{eq:electron-internal-energy} \\
    \mathbf{u}_i = \frac{1}{ne}\sum_sq_sn_s\mathbf{u}_s, \\
    n \equiv n_i = \frac{1}{e}\sum_{s}q_sn_s = n_e.
\end{align}

\begin{table}[h]
    \centering
    \begin{tabular}{|c|c|}
        \hline
        Symbol & Description \\
        \hline
         $s$ & Ion species \\
         $f_s$ & Probability Distribution Function for Ion Species $s$ \\
         $\mathbf{v}$ & Velocity Coordinate \\
         $\mathbf{E}$ & Electric Field \\
         $\mathbf{E}^*$ & Electric Field without Collisional Friction $\eta \mathbf{J}$ \\
         $\mathbf{B}$ & Magnetic Field \\
         $\mathbf{J}$ & Current Density, $\mathbf{J} = \frac{1}{\mu_0}\nabla\times\mathbf{B}$ \\
         $\mathbf{u}_i$ & Ion Current Carrying Drift Velocity \\
         $\mathbf{u}_e$ & Bulk Electron Velocity, $\mathbf{u}_e = \mathbf{u}_i - \frac{1}{ne}\mathbf{J}$ \\
         $\mathbf{q}_e$ & Electron Heat Flux, $\mathbf{q}_e = -\kappa_e\nabla(p_e/n)$ \\
         $\overset\leftrightarrow{\Pi}_e$ & Electron Viscous Tensor, $\overset\leftrightarrow{\Pi}_e = -\mu_e\nabla\mathbf{u}_e$ \\
         $n$ & Quasi-neutral Number Density \\
         $H_e$ & Ohmic and Electron Viscous Heating, $H_e = \eta\lVert\mathbf{J}\rVert^2 - \overset\leftrightarrow{\Pi}_e : \nabla\mathbf{u}_e$ \\
         $p_e$ & Electron Pressure \\
         $T_e$ & Electron Temperature \\
         $n_e$ & Electron Number Density \\
         $\mu_e$ & Electron Viscosity \\
         $\kappa_e$ & Electron Heat Conductivity \\
         $q_s$ & Charge of ion species $s$ \\
         $m_s$ & Mass of ion species $s$ \\
         $\eta$ & Resistivity \\
         $\gamma$ & Ratio of Specific Heats \\
         $\mu_0$ & Vacuum Permeability \\
         $e$ & Elementary Charge \\
        \hline
    \end{tabular}
    \caption{Continuous Hybrid Model Variable Lookup}
    \label{tab:model-symbol-lookup}
\end{table}

While we retain the electron viscous, heat flux, and Ohmic terms $\overset\leftrightarrow{\Pi}_e$, $\mathbf{q}_e$, $H_e$, and $\eta\mathbf{J}$ in the equations above for generality, the test problems considered in this work do not include them. Accordingly, for the remainder of the paper we set $\kappa_e = \mu_e = \eta = 0$, so that $\mathbf{E} = \mathbf{E}^*$ and these source terms drop out. Reintroducing them in either the continuous setting or the discrete form of \cite{stanier2019fully} does not alter the conservation arguments that follow.

Given a species $s$, we compute a number of key quantities by taking the moments of the distribution function $f_s$: number density, bulk velocity, and (kinetic) energy moments at space $\mathbf{x}$ and time $t$ by the following three formulae, respectively
    \begin{align}
        n_s(\mathbf{x},t) &\coloneqq \int_{\Omega_\mathbf{v}}f_s\left(\mathbf{x},\mathbf{v},t\right)d\mathbf{v}, \label{eq:mass-moment-continuous-s} \\
        \mathbf{u}_s(\mathbf{x},t) &\coloneqq \frac{1}{n_s}\int_{\Omega_\mathbf{v}}\mathbf{v}f_s\left(\mathbf{x},\mathbf{v},t\right)d\mathbf{v}, \label{eq:momentum-moment-continuous-s} \\
        \kappa_s(\mathbf{x},t) &\coloneqq \frac{1}{2}m_s\int_{\Omega_\mathbf{v}}\lVert\mathbf{v}\rVert^2f_sd\mathbf{v}. \label{eq:KE-moment-continuous-s}
    \end{align}

Additionally, the pressure tensor and energy flux vectors are defined as
\begin{align}
    \mathcal{P}_s(\mathbf{x},t) &\coloneqq \int_{\Omega_\mathbf{v}}m_s\mathbf{v}\otimes\mathbf{v}f_sd\mathbf{v}, \label{eq:pressure-dyad} \\
    \mathbf{Q}_s(\mathbf{x},t) &\coloneqq \frac{1}{2}\int_{\Omega_\mathbf{v}}m_s\lVert\mathbf{v}\rVert^2\mathbf{v}f_sd\mathbf{v}. \label{eq:heat-flux}
\end{align}

To reduce some tedium, for the remainder of the paper we introduce two common abuses of notation for vectors $\mathbf{a}$ and $\mathbf{b}$: $\mathbf{a}\mathbf{b} \equiv \mathbf{a}\otimes\mathbf{b}$ and $\lVert\mathbf{a}\rVert \equiv a$.

\subsection{Conservation Properties of the Continuous Hybrid Model} \label{subsection:conservation-properties-continuum}

In this section we discuss the conservation of mass, momentum, and energy. We find the contributions to these three values by taking moments of the Vlasov equation \eqref{eq:mass-moment-continuous-s}-\eqref{eq:KE-moment-continuous-s} and combining them with the electron fluid equations. Taking the moments, we will find the following identity useful:

\begin{equation} \label{eq:IBP-velocity}
    \int\chi\left(\mathbf{F}\cdot\nabla_v f_s\right)d\mathbf{v} = \cancel{\int_{\Omega_v}\nabla_v\cdot\left(\mathbf{F}\chi f_s\right)d\mathbf{v}} - \int_{\Omega_v}f_s\mathbf{F}\cdot\nabla_v\chi d\mathbf{v} - \cancel{\int_{\Omega_v}f_s\chi\nabla_v\cdot\mathbf{F}d\mathbf{v}} = -\int_{\Omega_v}f_s\mathbf{F}\cdot\nabla_v\chi d\mathbf{v}.
\end{equation}

The first term disappears assuming $f_s \rightarrow 0$ as $v \rightarrow \infty$, while the third disappears due to the velocity divergence of the Lorentz force $\mathbf{F}$ being zero. Here $\chi(\mathbf{v})$ is some physical property of the particles; for our purposes it is either $1$, $m_s\mathbf{v}$, or $\frac{1}{2}m_s v^2$, yielding mass, momentum, and energy, respectively. So the mass, momentum, and energy moments of the $v$-advection component of the Vlasov equation are as follows:
\begin{align}
    \int_{\Omega_v}\mathbf{F}\cdot\nabla_vf_s d\mathbf{v}  &= 0, \label{eq:IBP-mass} \\
    \int_{\Omega_v}m_s\mathbf{v}\mathbf{F}\cdot\nabla_vf_s  d\mathbf{v} &= -q_sn_s\left(\mathbf{E}^* + \mathbf{u}_s\times\mathbf{B}\right), \label{eq:IBP-momentum} \\
    \int_{\Omega_v}\frac{1}{2}m_sv^2\mathbf{F}\cdot\nabla_vf_s  d\mathbf{v} &= -q_sn_s\mathbf{E}^*\cdot\mathbf{u}_s. \label{eq:IBP-energy}
\end{align}

These form the right-hand side (RHS) of the Vlasov components of the continuity equations for mass, momentum, and energy. The first may be found by either integration by parts (IBP) or the divergence theorem, the latter two appeal to IBP. In a continuous (non-discrete) system there are no boundaries in velocity space. The boundaries appear when we discretize on a mesh, which may cause errors with the above identities. The workaround for these issues is discussed in Appendix \ref{appendix:lagrange-multipliers}.

\subsubsection{Conservation of Mass} \label{section:conservation-mass-continuum}

Taking the zeroth moment \eqref{eq:mass-moment-continuous-s} of the Vlasov equation \eqref{eq:vlasov-s} yields the continuity equation of number density for a species $s$:
\begin{equation}
    \frac{\partial n_s}{\partial t} + \nabla\cdot\left(n_s\mathbf{u}_s\right) = 0.
\end{equation}

The RHS comes from \eqref{eq:IBP-mass}. Summing over all $s$ gives us our continuity equation of number density (see Table \ref{tab:model-symbol-lookup} for variable details).
\begin{equation} \label{eq:local-mass-conservation-continuous}
    \frac{\partial n}{\partial t} + \nabla\cdot\left(n \mathbf{u}_i\right) = 0.
\end{equation}

Assuming appropriate boundary conditions (BCs) in physical space, such as periodic, mass density will remain globally constant, as integrating over a spatial derivative in a periodic setting goes to zero.

\subsubsection{Conservation of Momentum}

If we take the first moment \eqref{eq:momentum-moment-continuous-s}, weighted by species mass $m_s$, of the Vlasov equation \eqref{eq:vlasov-s} we get the momentum density equation for an ion species $s$:
\begin{equation}
    \frac{\partial}{\partial t}\left[m_sn_s\mathbf{u}_s\right] + \nabla\cdot\mathcal{P}_s = q_sn_s\left(\mathbf{E}^* + \mathbf{u}_s\times\mathbf{B}\right).
\end{equation}

The RHS comes from \eqref{eq:IBP-momentum}. Recall $\mathcal{P}_s$ is the pressure dyad \eqref{eq:pressure-dyad}. Again summing over the species $s$, and recalling $\sum_sq_sn_s = en$ and $\sum_sq_sn_s\mathbf{u}_s = en\mathbf{u}_i$ we get
\begin{equation}
    \sum_s\left(\frac{\partial}{\partial t}\left[m_sn_s\mathbf{u}_s\right] + \nabla\cdot\mathcal{P}_s\right) = en\left(\mathbf{E}^* + \mathbf{u}_i\times\mathbf{B}\right).
\end{equation}

Recalling the definition of $\mathbf{E}^*$ \eqref{eq:E-definition}, we employ the identity $\left(\nabla\times\mathbf{B}\right)\times\mathbf{B} = \nabla\cdot\left(\mathbf{B}\mathbf{B} - \frac{1}{2}B^2\mathcal{I}\right)$, where $\mathcal{I}$ is the identity matrix, and see
\begin{equation} \label{eq:local-momentum-density-conservation-continuous}
    \sum_s\frac{\partial}{\partial t}\left[m_sn_s\mathbf{u}_s\right] + \nabla\cdot\left[\sum_s\mathcal{P}_s + \left(p_e + \frac{1}{2}\frac{1}{\mu_0}B^2\right)\mathcal{I} - \frac{1}{\mu_0}\mathbf{B}\mathbf{B}\right] = 0.
\end{equation}

The time derivative of the total momentum over the whole domain will only vanish given suitable boundary conditions such as periodic.

\subsubsection{Conservation of Energy}

Taking the (contracted) second moment \eqref{eq:KE-moment-continuous-s} of the Vlasov equation \eqref{eq:vlasov-s} yields the continuity equation for kinetic energy
\begin{equation}
    \frac{\partial\kappa_s}{\partial t} + \nabla\cdot\mathbf{Q}_s = q_sn_s\mathbf{E}^*\cdot\mathbf{u}_s.
\end{equation}

The RHS comes from \eqref{eq:IBP-energy}. Recall $\mathbf{Q}_s$ is the energy flux vector \eqref{eq:heat-flux}. Once more summing over $s$ and employing the definition of $\mathbf{u}_i$ we get
\begin{equation} \label{eq:summed-KE-continuity}
    \sum_s\left(\frac{\partial\kappa_s}{\partial t} + \nabla\cdot\mathbf{Q}_s\right) = en\mathbf{E}^*\cdot\mathbf{u}_i.
\end{equation}

We take the electron internal energy equation \eqref{eq:electron-internal-energy} and rearrange for reasons that will be clear presently:
\begin{equation} \label{eq:electron-internal-energy-rearrange}
    (\gamma-1)^{-1}\frac{\partial p_e}{\partial t} + \nabla\cdot\left(\gamma(\gamma-1)^{-1}\left(\mathbf{u}_ep_e\right)\right) = \mathbf{u}_e\cdot\nabla p_e.
\end{equation}

We know $\frac{\partial}{\partial t}\left[\mathbf{B}\cdot\mathbf{B} \equiv B^2\right] = 2\mathbf{B}\cdot\frac{\partial\mathbf{B}}{\partial t}$. We combine this with the definition of $\mathbf{B}$ \eqref{eq:curl-E} to see 
\begin{equation} \label{eq:magnetic-energy}
    \frac{1}{2}\frac{1}{\mu_0}\frac{\partial B^2}{\partial t} = -\frac{1}{\mu_0}\mathbf{B}\cdot\left(\nabla\times\mathbf{E}\right).
\end{equation}

The identity $\nabla\cdot\left(\mathbf{A}\times\mathbf{B}\right) = \mathbf{B}\cdot\left(\nabla\times\mathbf{A}\right) - \mathbf{A}\cdot\left(\nabla\times\mathbf{B}\right)$, combined with the definition of $\mathbf{J}$ (Table \ref{tab:model-symbol-lookup}), turns \eqref{eq:magnetic-energy} into
\begin{equation} \label{eq:poynting-without-electric-displacement}
    \frac{1}{2}\frac{1}{\mu_0}\frac{\partial B^2}{\partial t} + \frac{1}{\mu_0}\nabla\cdot\left(\mathbf{E}\times\mathbf{B}\right) = -\mathbf{J}\cdot\mathbf{E}.
\end{equation}

Combining the summed ion kinetic energy continuity equation \eqref{eq:summed-KE-continuity}, the electron pressure evolution \eqref{eq:electron-internal-energy-rearrange}, and Poynting's theorem \eqref{eq:poynting-without-electric-displacement}, we get the total energy equation
\begin{equation} \label{eq:total-energy-conservation-continuous-penultimate}
    \frac{\partial}{\partial t}\left[\sum_s\kappa_s + \left(\gamma-1\right)^{-1}p_e + \frac{1}{2}\frac{1}{\mu_0}B^2\right] + \nabla\cdot\left[\sum_s\mathbf{Q}_s + \gamma\left(\gamma-1\right)^{-1}\mathbf{u}_ep_e + \frac{1}{\mu_0}\mathbf{E}\times\mathbf{B}\right] = RHS.
\end{equation}

We consider the RHS with $\eta=0$ to see

\begin{align}
    \begin{split}
        RHS &=en\mathbf{E}^*\cdot\mathbf{u}_i - \mathbf{J}\cdot\mathbf{E}^* + \mathbf{u}_e\cdot\nabla p_e \\
        &= en\left(-\mathbf{u}_i\times\mathbf{B} + \frac{1}{ne}\mathbf{J}\times\mathbf{B} - \frac{1}{ne}\nabla p_e\right)\cdot\mathbf{u}_i - \mathbf{J}\cdot\left(-\mathbf{u}_i\times\mathbf{B} + \frac{1}{ne}\mathbf{J}\times\mathbf{B} - \frac{1}{ne}\nabla p_e\right) + \mathbf{u}_e\cdot\nabla p_e \\
        &= (\mathbf{J}\times\mathbf{B})\cdot\mathbf{u}_i - \nabla p_e\cdot\mathbf{u}_i + \mathbf{J}\cdot\left(\mathbf{u}_i\times\mathbf{B}\right) + \frac{1}{ne}\mathbf{J}\cdot\nabla p_e + \mathbf{u}_e\cdot\nabla p_e \\
        &= \left(\mathbf{u}_e - \mathbf{u}_i + \frac{1}{ne}\mathbf{J}\right)\cdot\nabla p_e + \left(\left(\mathbf{J}\times\mathbf{B}\right)\cdot\mathbf{u}_i + \mathbf{J}\cdot\left(\mathbf{u}_i \times \mathbf{B}\right)\right).
    \end{split}
\end{align}

The first term is zero by definition of $\mathbf{J} = en\left(\mathbf{u}_i - \mathbf{u}_e\right)$. The second is zero by a vector identity, $\mathbf{A}\cdot\left(\mathbf{B}\times\mathbf{C}\right) = \mathbf{B}\cdot\left(\mathbf{C}\times\mathbf{A}\right)$. Thus $RHS = 0$, and our total energy equation takes its final form:

\begin{equation} \label{eq:total-energy-conservation-continuous}
    \frac{\partial}{\partial t}\left[\sum_s\kappa_s + \left(\gamma-1\right)^{-1}p_e + \frac{1}{2}\frac{1}{\mu_0}B^2\right] + \nabla\cdot\left[\sum_s\mathbf{Q}_s + \gamma\left(\gamma-1\right)^{-1}\mathbf{u}_ep_e + \frac{1}{\mu_0}\mathbf{E}\times\mathbf{B}\right] = 0.
\end{equation}

As with our momentum equation, the spatial derivative will go to zero assuming certain boundary conditions (e.g. periodic), which ensures a constant total energy.
\subsection{The Discrete Hybrid Model} \label{subsection:discrete-hybrid-model}

We have discretized our phase space mesh as follows
\begin{align}
    \boldsymbol{\eta} &= \left\{\eta_{min} + \Delta \eta \left(i - \frac{1}{2}\right)\right\}_{i=1}^{N_{\eta}}, \quad \boldsymbol{\eta} \in \left\{\mathbf{x}, \mathbf{y}, \mathbf{z}\right\}. \\
    \mathbf{v}_\eta &= \left\{v_{\eta,min} + \Delta v_\eta \left(j - \frac{1}{2}\right)\right\}_{j=1}^{N_{v_\eta}}, \quad \eta \in \left\{x,y,z\right\}, v_{\eta,min} = -v_{\eta,max}.
\end{align}

The discrete phase space is then a tensor product of six vectors:
\begin{equation}
    \Omega = \Omega_{\mathbf{x}} \otimes \Omega_\mathbf{v} = \bigotimes_{\boldsymbol{\eta}\in\{\mathbf{x},\mathbf{y},\mathbf{z}\}}\boldsymbol{\eta} \otimes \bigotimes_{\eta\in\{x,y,z\}}\mathbf{v}_\eta.
\end{equation}

Some care is needed here, as the meaning of $\mathbf{v}$ shifts upon discretization. In the continuous model $\mathbf{v} = (v_x, v_y, v_z)$ is a single point of velocity space, each component a scalar. Discretely, each component $v_\eta$ is replaced by an entire grid $\mathbf{v}_\eta$ of $N_{v_\eta}$ values, so $\mathbf{v}$ becomes a triple of vectors rather than a triple of scalars, i.e. a vector of vectors. A quantity depending on a single velocity direction, such as the coordinate $v_x$, lives in the full velocity space $\Omega_\mathbf{v} = \mathbf{v}_x \otimes \mathbf{v}_y \otimes \mathbf{v}_z$ only after we embed its grid by tensoring with the all-ones vectors $\mathbf{1}_{v_x}, \mathbf{1}_{v_y}, \mathbf{1}_{v_z}$ in the remaining directions:
\begin{equation} \label{eq:velocity-embedding}
    v_x \mapsto \mathbf{v}_x \otimes \mathbf{1}_{v_y} \otimes \mathbf{1}_{v_z}, \quad
    v_y \mapsto \mathbf{1}_{v_x} \otimes \mathbf{v}_y \otimes \mathbf{1}_{v_z}, \quad
    v_z \mapsto \mathbf{1}_{v_x} \otimes \mathbf{1}_{v_y} \otimes \mathbf{v}_z.
\end{equation}

Thus the discrete mass weight becomes the fully embedded all-ones tensor $1 \coloneqq \mathbf{1}_{v_x}\otimes\mathbf{1}_{v_y}\otimes\mathbf{1}_{v_z}$, the discrete momentum weight $\mathbf{v}$ becomes the triple of embedded grids above, and the discrete energy weight $v^2$ becomes the sum of their squares, $\mathbf{v}_x^2 \otimes \mathbf{1}_{v_y} \otimes \mathbf{1}_{v_z} + \mathbf{1}_{v_x} \otimes \mathbf{v}_y^2 \otimes \mathbf{1}_{v_z} + \mathbf{1}_{v_x} \otimes \mathbf{1}_{v_y} \otimes \mathbf{v}_z^2$, where each $\mathbf{v}_\eta^2 \coloneqq \mathbf{v}_\eta \star \mathbf{v}_\eta$ is the Hadamard square of the one-dimensional velocity grid. This same embedding underlies the moment basis vectors $\mathbf{V}_i$ of Section \ref{subsection:conserving-truncation}.

Once again setting $\kappa_e = \mu_e = \eta = 0$, we discretize the hybrid fluid-electron kinetic-ion system \eqref{eq:vlasov-s}-\eqref{eq:electron-internal-energy} using the implicit midpoint scheme for time. Spatial derivatives are computed using central differences or the Quadratic Upstream Interpolation for Convective Kinematics (QUICK) method \cite{Leonard_1979}. QUICK is only used to compute the spatial advection term for the Vlasov solve. Velocity derivatives are computed using central difference. For a value $\chi$ at known integer timesteps $n$ and $n+1$, we denote its midpoint value as $\chi^{n+\frac{1}{2}} \equiv \frac{1}{2}\left(\chi^n + \chi^{n+1}\right)$. While our Vlasov solver still considers $\mathbf{F}$ in terms of $\mathbf{E}$ and $\mathbf{B}$, the existing fluid framework we make use of (the same as \cite{stanier2019fully}) utilizes a scalar and vector potential ($\phi$ and $\mathbf{A}$, respectively) formulation with the Weyl gauge ($\phi = 0$), yielding $\mathbf{B} = \mathbf{B}_0 + \nabla\times\mathbf{A}$ and $\mathbf{E} = -\frac{\partial\mathbf{A}}{\partial t}$, where $\mathbf{B}_0$ is a spatially uniform, static background field and $\mathbf{A}$ carries the field perturbation. The uniform $\mathbf{B}_0$ is retained separately rather than absorbed into $\mathbf{A}$; on a periodic domain, such as those used in our test problems, this split is in fact required, since a single-valued periodic vector potential cannot represent a nonzero mean field. As $\mathbf{B}_0$ is constant it drops out of every spatial derivative below, so $\nabla\times\mathbf{A}$ and $\nabla\times\left(\nabla\times\mathbf{A}\right)$ may be read as the corresponding operators acting on the total field; in particular $\mathbf{J} = \frac{1}{\mu_0}\nabla\times\mathbf{B} = \frac{1}{\mu_0}\nabla\times\left(\nabla\times\mathbf{A}\right)$ is unchanged. This gives us the following:

\begin{align}
    -\frac{\mathbf{A}_i^{n+1} - \mathbf{A}_i^n}{\Delta t} &= \mathbf{E}_i^{*,n+\frac{1}{2}}. \label{eq:discrete-E-dA/dt} \\
    \mathbf{E}_i^{*,n+\frac{1}{2}} &= -\mathbf{u}_{i}^{n+\frac{1}{2}}\times\left(\nabla\times\mathbf{A}\right)_i^{n+\frac{1}{2}} + \frac{1}{en_i^{n+\frac{1}{2}}}\left(\mathbf{J}\times\left(\nabla\times\mathbf{A}\right) - \nabla p_e\right)_i^{n+\frac{1}{2}} \label{eq:discrete-E} \\
    \begin{split}
        \frac{p_{e,i}^{n+1} - p_{e,i}^{n}}{\Delta t} + \gamma\left(\nabla\cdot\left[\mathbf{u}_ep_e\right]\right)_i^{n+\frac{1}{2}} &= \left(\gamma-1\right)\mathbf{u}_{e,i}^{n+\frac{1}{2}}\cdot\left(\nabla p_e\right)_i^{n+\frac{1}{2}}, \label{eq:IM-electron-pressure}
    \end{split} \\
    \mathbf{J}_i^{n+\frac{1}{2}} &= \frac{1}{\mu_0}\nabla\times\left(\nabla\times\mathbf{A}^{n+\frac{1}{2}}\right)_i. \label{eq:J-discrete-curl-curl-A}
\end{align}

Integration will now take the form of a finite summation with multi-index $j$:
\begin{equation} \label{eq:discrete-integral}
    \left\langle\mathbf{a},\mathbf{b}\right\rangle = \Delta v_x\Delta v_y\Delta v_z\sum_{j}a_jb_j \equiv \Delta V\sum_{j}a_jb_j.
\end{equation}

This gives the following discrete moment identities:
\begin{align}
    \left\langle \mathbf{1},\mathbf{F}\cdot\nabla_vf_s\right\rangle &= 0, \label{eq:IBP-mass-discrete} \\
    \left\langle m_s\mathbf{v},\mathbf{F}\cdot\nabla_vf_s\right\rangle &= -q_sn_s\left(\mathbf{E}^* + \mathbf{u}_s\times\mathbf{B}\right), \label{eq:IBP-momentum-discrete} \\
    \left\langle \frac{1}{2}m_sv^2,\mathbf{F}\cdot\nabla_vf_s\right\rangle &= -q_sn_s\mathbf{E}^*\cdot\mathbf{u}_s. \label{eq:IBP-energy-discrete}
\end{align}

There is a caveat to note: these identities use IBP and assume $\lim_{\mathbf{v} \rightarrow \{\mathbf{v}_{min}, \mathbf{v}_{max}\}}f = 0$. If this is not the case (i.e. $f$ has nonzero values at the boundaries) there may be a remainder that comes with IBP. For dealing with these remainders, see Appendix \ref{appendix:lagrange-multipliers}.
\subsection{A Conservative Update Scheme} \label{section:conservative-update-scheme}

We now prove the implicit midpoint update conserves the first three moments of the distribution function \eqref{eq:mass-moment-continuous-s}-\eqref{eq:KE-moment-continuous-s} by satisfying the discrete analogues of \eqref{eq:local-mass-conservation-continuous}, \eqref{eq:local-momentum-density-conservation-continuous}, and \eqref{eq:total-energy-conservation-continuous}.

We have the collisionless Vlasov equation \eqref{eq:vlasov-s}, which we semi-discretize using the implicit midpoint method
\begin{equation}
    \frac{f_{s,i}^{n+1} - f_{s,i}^{n}}{\Delta t} = -\frac{1}{2}\left[\mathbf{v}\cdot\nabla_x\left(f_{s,i}^{n+1} + f_{s,i}^{n}\right) + \mathbf{F}_i^{n+\frac{1}{2}}\cdot\nabla_v\left(f_{s,i}^{n+1} + f_{s,i}^n\right)\right].
\end{equation}

The following proofs will be in $1$D$3$V.\footnote{The only difference in extending to $3$D$3$V would be in the spatial derivative component, which will always disappear due to the conservative discretization of the spatial flux terms (central difference or QUICK). We reduce to 1 dimension to save some tedium.} We can compute the spatial difference a number of ways (see Section \ref{section:implementation} for details), so we write it as a divergence of flux terms:
\begin{equation}
    \frac{f_{s,i}^{n+1} - f_{s,i}^{n}}{\Delta t} = -\frac{1}{2}\left[\frac{\left(\Phi^{n+1}_s + \Phi^{n}_s\right)_{i+\frac{1}{2}} - \left(\Phi^{n+1}_s + \Phi^{n}_s\right)_{i-\frac{1}{2}}}{\Delta x} + \mathbf{F}_i^{n+\frac{1}{2}}\cdot\nabla_v\left(f^{n+1}_{s,i} + f^n_{s,i}\right)\right].
\end{equation}

Here $\Phi$ is the flux across the cell interface. We consider the term in the brackets for the following analysis; if we can show the inner product of it and $\chi(\mathbf{v})$ to be zero, we have proven that the time derivative of that quantity is zero, i.e. it is conserved.

\subsubsection{Conservation of Mass}
We take the zeroth moment of the RHS:
\begin{align}
    \left\langle \mathbf{1}, \frac{\left(\Phi^{n+1}_s + \Phi^{n}_s\right)_{i+\frac{1}{2}} - \left(\Phi^{n+1}_s + \Phi^{n}_s\right)_{i-\frac{1}{2}}}{\Delta x} + \mathbf{F}_i^{n+\frac{1}{2}}\cdot\nabla_v\left(f^{n+1}_{s,i} + f^n_{s,i}\right)\right\rangle
\end{align}

The first component will sum to zero over certain boundary conditions (e.g. periodic or reflecting). We now consider the second term of the RHS:
\begin{equation}
    \langle\mathbf{1},\mathbf{F}_i^{n+\frac{1}{2}}\cdot\nabla_vf_{s,i}^{n}\rangle = \langle\mathbf{1},\mathbf{F}_i^{n+\frac{1}{2}}\cdot\nabla_vf_{s,i}^{n+1}\rangle = 0.
\end{equation}

We know this holds in an infinitely long velocity space domain, but we can also ensure it in a finite velocity domain using zero-flux antisymmetric boundary conditions (see Appendix \ref{subsection:lagrange-multipliers-mass}).

\subsubsection{Conservation of Momentum}

We take the first moment and sum over species from the outset. Unlike mass, which is conserved per species, momentum is exchanged among species through the fields, so only the total ion momentum is conserved. Quasi-neutrality supplies the identities $\sum_s q_s n_{s,i} = e n_i$ and $\sum_s q_s n_{s,i}\mathbf{u}_{s,i} = e n_i\mathbf{u}_i$, where $n_i$ and $\mathbf{u}_i$ are the total ion density and flow velocity. As before, we disregard the first term of the RHS, assuming appropriate boundary conditions sum it to zero, and turn to the second term:
\begin{align}
    \begin{split}
        \sum_s\left\langle m_s\mathbf{v}, \mathbf{F}_{s,i}^{n+\frac{1}{2}}\cdot\nabla_v\left(f_{s,i}^{n+1} + f_{s,i}^{n}\right)\right\rangle_i &= -\sum_s q_s\left(n_{s,i}^{n+1} + n_{s,i}^{n}\right)\mathbf{E}_{i}^{*,n+\frac{1}{2}} - \sum_s q_s\left(n_{s,i}^{n+1}\mathbf{u}_{s,i}^{n+1} + n_{s,i}^{n}\mathbf{u}_{s,i}^{n}\right)\times\left(\nabla\times\mathbf{A}\right)_{i}^{n+\frac{1}{2}} \\
        &= -e\left(n_{i}^{n+1} + n_{i}^n\right)\left(-\mathbf{u}_{i}\times\left(\nabla\times\mathbf{A}\right)_i^{n+\frac{1}{2}} + \frac{1}{en^{n+\frac{1}{2}}_{i}}\left[\mathbf{J}\times\left(\nabla\times\mathbf{A}\right) - \nabla p_e\right]^{n+\frac{1}{2}}_i\right) \\
        &  -e\left(n_{i}^{n+1}\mathbf{u}_{i}^{n+1} + n_{i}^{n}\mathbf{u}_{i}^{n}\right)\times\left(\nabla\times\mathbf{A}\right)_{i}^{n+\frac{1}{2}}.
    \end{split}
\end{align}

Again note that for a finite velocity space domain Eq. \eqref{eq:IBP-momentum-discrete} may have a remainder term due to the boundary contribution when $f$ remains finite (see Appendix \ref{subsection:lagrange-multipliers-momentum}).

We note
\begin{equation}
    -e\left(n_{i}^{n+1} + n_{i}^{n}\right)\left(-\mathbf{u}_{i}^{n+\frac{1}{2}}\times\left(\nabla\times\mathbf{A}\right)_i^{n+\frac{1}{2}}\right) = 2en_i^{n+\frac{1}{2}}\mathbf{u}_{i}^{n+\frac{1}{2}}\times\left(\nabla\times\mathbf{A}\right)_i^{n+\frac{1}{2}}.
\end{equation}

We also note
\begin{equation}
    - e\left(n_{i}^{n+1}\mathbf{u}_{i}^{n+1} + n_{i}^{n}\mathbf{u}_{i}^{n}\right)\times\left(\nabla\times\mathbf{A}\right)_{i}^{n+\frac{1}{2}} = -2e\left(n_{i}\mathbf{u}_{i}\right)^{n+\frac{1}{2}}\times\left(\nabla\times\mathbf{A}\right)_i^{n+\frac{1}{2}}.
\end{equation}

Given our definition of $\mathbf{u}_{i}^{n+\frac{1}{2}} \coloneqq \left(n\mathbf{u}\right)_i^{n+\frac{1}{2}} / n_{i}^{n+\frac{1}{2}}$, it follows $\left(n_{i}\mathbf{u}_{i}\right)^{n+\frac{1}{2}} = n_{i}^{n+\frac{1}{2}}\mathbf{u}_{i}^{n+\frac{1}{2}}$, and so these two terms cancel.

Recalling $\mathbf{J} = \frac{1}{\mu_0}\nabla\times\mathbf{B} = \frac{1}{\mu_0}\nabla\times\left(\nabla\times\mathbf{A}\right)$, in like manner to \cite{stanier2019fully}, through some discrete IBP and some rearranging, we ultimately get the following:
\begin{equation}
    \sum_i\left(\left(\nabla\times\left(\nabla\times\mathbf{A}\right)\right)\times\left(\nabla\times\mathbf{A}\right)\right)_i = \sum_i\left(\nabla\cdot\left(\left(\nabla\times\mathbf{A}\right)\left(\nabla\times\mathbf{A}\right)\right)\right)_i -\sum_i\left(\left(\nabla\times\mathbf{A}\right)\left(\nabla\cdot\left(\nabla\times\mathbf{A}\right)\right)\right)_i - \frac{1}{2}\sum_i\left(\nabla\lVert\nabla\times\mathbf{A}\rVert^2\right)_i.
\end{equation}

The middle term vanishes so long as the discrete $\nabla$ operator preserves the $\nabla\cdot\left(\nabla\times\mathbf{v}\right) = 0, \, \forall \, \mathbf{v}$ identity (we employ central differencing, so this holds pointwise and requires no boundary conditions), leaving the dyad-divergence term $\sum_i\left(\nabla\cdot\left(\left(\nabla\times\mathbf{A}\right)\left(\nabla\times\mathbf{A}\right)\right)\right)_i$ and $-\frac{1}{2}\sum_i\left(\nabla\lVert\nabla\times\mathbf{A}\rVert^2\right)_i$. These are a discrete divergence and a discrete gradient, respectively, so each telescopes and sums to zero, assuming boundary conditions such as periodic.

\subsubsection{Conservation of Energy} \label{subsubsection:energy-conservation-discrete}

We take the contracted second moment, again summing over species. Once more the first term sums to zero and we turn to consider the second term:
\begin{align}
    \begin{split}
        \sum_s\left\langle m_s\frac{1}{2}v^2, \mathbf{F}_{s,i}^{n+\frac{1}{2}}\cdot\nabla_vf_{s,i}\right\rangle_{i} &= -\sum_s n_{s,i}^{n+\frac{1}{2}}q_s\mathbf{u}_{s,i}^{n+\frac{1}{2}}\cdot\mathbf{E}_{i}^{*,n+\frac{1}{2}} \\
        &= -n_i^{n+\frac{1}{2}}e\mathbf{u}_i^{n+\frac{1}{2}}\cdot\left(-\mathbf{u}_{i}^{n+\frac{1}{2}}\times\left(\nabla\times\mathbf{A}\right)_i^{n+\frac{1}{2}} + \frac{1}{en_i^{n+\frac{1}{2}}}\left[\mathbf{J}\times\left(\nabla\times\mathbf{A}\right) - \nabla p_e\right]_i^{n+\frac{1}{2}}\right) \\
        &= -\mathbf{u}_i^{n+\frac{1}{2}}\cdot\left(\mathbf{J}\times\left(\nabla\times\mathbf{A}\right)\right)_i^{n+\frac{1}{2}} + \mathbf{u}_i^{n+\frac{1}{2}}\cdot(\nabla p_e)_i^{n+\frac{1}{2}}
    \end{split}
\end{align}

We appealed to \eqref{eq:IBP-energy-discrete} for this to work. With this identity may come a remainder term from the boundary contribution to the integration by parts. See Appendix \ref{subsection:lagrange-multipliers-energy}. We define $K_{ion}$ as the sum total of kinetic-ion energy, found by integrating over space, which in the discrete setting requires summing over the spatial cells multiplied by their volume $\Delta X \coloneqq \prod_{\eta\in\{x,y,z\}}\Delta \eta$ (for the single-dimensional case, we only consider $\Delta X = \Delta x$). Noting the change in $K_{ion}$ over a timestep is equal to the negative of the above, we see our total kinetic-ion energy update equation
\begin{equation}
    \frac{K_{ion}^{n+1} - K_{ion}^{n}}{\Delta t} = \sum_i\Delta X\left[\mathbf{u}_i^{n+\frac{1}{2}}\cdot\left(\mathbf{J}\times\left(\nabla\times\mathbf{A}\right)\right)_i^{n+\frac{1}{2}} - \mathbf{u}_i^{n+\frac{1}{2}}\cdot(\nabla p_e)_i^{n+\frac{1}{2}}\right].
\end{equation}

The implicit midpoint discretization of the electron pressure equation is found as follows
\begin{equation} 
    \frac{p_{e,i}^{n+1} - p_{e,i}^{n}}{\Delta t} + \gamma\left(\nabla\cdot\left[\mathbf{u}_ep_e\right]\right)_i^{n+\frac{1}{2}} = \left(\gamma-1\right)\mathbf{u}_{e,i}^{n+\frac{1}{2}}\cdot\left(\nabla p_e\right)_i^{n+\frac{1}{2}}.
\end{equation}

Discrete electron energy at time level $n$ may likewise be found by integrating over space:
\begin{equation}
    K_{ele}^n = \sum_{i}\Delta X(\gamma-1)^{-1}p_{e,i}^n.
\end{equation}

Plugging this into \eqref{eq:IM-electron-pressure}, we see
\begin{align}
    \begin{split}
        \frac{K_{ele}^{n+1} - K_{ele}^{n}}{\Delta t} &= \sum_i\frac{\Delta X}{\gamma-1}\frac{p_{e,i}^{n+1} - p_{e,i}^{n}}{\Delta t} \\
        &= \sum_i\frac{\Delta X}{\gamma-1}\left[-\gamma\left(\nabla\cdot\left[\mathbf{u}_ep_e\right]\right)_i^{n+\frac{1}{2}} + \left(\gamma-1\right)\mathbf{u}_{e,i}^{n+\frac{1}{2}}\cdot\left(\nabla p_e\right)_i^{n+\frac{1}{2}}\right] \\
        &= \sum_i\Delta X\left[-\frac{\gamma}{\gamma-1}\left(\nabla\cdot\left[\mathbf{u}_ep_e\right]\right)_i^{n+\frac{1}{2}} + \mathbf{u}_{e,i}^{n+\frac{1}{2}}\cdot\left(\nabla p_e\right)_i^{n+\frac{1}{2}}\right].
    \end{split}
\end{align}

The sum of the magnetic energy at cell centers at time level $n$ is
\begin{equation}
    W_B^n = \sum_i \Delta X\frac{1}{2}\frac{1}{\mu_0}\left[(\nabla\times\mathbf{A})_i^n\right]^2
\end{equation}

So we see the discrete time derivative of $W_B$ is
\begin{align}
    \begin{split}
        \frac{W_B^{n+1} - W_B^{n}}{\Delta t} &= \sum_i\Delta X\frac{1}{2}\frac{1}{\mu_0}\frac{\left[(\nabla\times\mathbf{A})_i^{n+1}\right]^2 - \left[(\nabla\times\mathbf{A})_i^{n}\right]^2}{\Delta t} \\
        &= \sum_i\Delta X\frac{1}{2}\frac{1}{\mu_0}\frac{\left[(\nabla\times\mathbf{A})_i^{n+1} + (\nabla\times\mathbf{A})_i^{n}\right] \cdot \left[(\nabla\times\mathbf{A})_i^{n+1} - (\nabla\times\mathbf{A})_i^{n}\right]}{\Delta t} \\
        &= \sum_i\Delta X\frac{1}{\mu_0}(\nabla\times\mathbf{A}_i^{n+\frac{1}{2}}) \cdot \left(\nabla\times\frac{\mathbf{A}_i^{n+1} - \mathbf{A}_i^{n}}{\Delta t}\right).
    \end{split}
\end{align}

We recall the identity $\left(\nabla\times\mathbf{a}\right)\cdot\left(\nabla\times\mathbf{b}\right) = \mathbf{b}\cdot\left(\nabla\times\nabla\times\mathbf{a}\right) - \nabla\cdot\left((\nabla\times\mathbf{a})\times\mathbf{b}\right)$\footnote{Pointwise this identity may have a remainder in the discrete setting, but assuming a periodic domain and central differencing, whose skew-symmetric stencil makes the discrete curl self-adjoint, the remainder vanishes over summation.} to write
\begin{align}
    \begin{split}
        \sum_i\Delta X\frac{1}{\mu_0}\left[\frac{\mathbf{A}_i^{n+1} - \mathbf{A}_i^{n}}{\Delta t}\cdot(\nabla\times\nabla\times\mathbf{A}_i^{n+\frac{1}{2}}) - \nabla\cdot\left((\nabla\times\mathbf{A}_i^{n+\frac{1}{2}})\times\frac{\mathbf{A}_i^{n+1} - \mathbf{A}_i^{n}}{\Delta t}\right)\right] = \sum_i\Delta X\left(-\mathbf{E}_i^{*,n+\frac{1}{2}}\right)\cdot\mathbf{J}_i^{n+\frac{1}{2}}.
    \end{split}
\end{align}

The second term goes to zero under appropriate boundary conditions. We applied \eqref{eq:discrete-E-dA/dt} and \eqref{eq:J-discrete-curl-curl-A} to acquire the $\mathbf{E}^{*}\cdot\mathbf{J}$ term. We then apply \eqref{eq:discrete-E} to finally acquire
\begin{equation}
    \sum_i\Delta X\left(\left(\mathbf{u}_{i}^{n+\frac{1}{2}}\times\left(\nabla\times\mathbf{A}\right)_i^{n+\frac{1}{2}}\right)\cdot\mathbf{J}_i^{n+\frac{1}{2}} + \frac{1}{en_i^{n+\frac{1}{2}}}\left(\nabla p_e\right)_i^{n+\frac{1}{2}}\cdot\mathbf{J}_i^{n+\frac{1}{2}}\right).
\end{equation}

We finally conclude $\frac{(K_{ion} + K_{ele} + W_B)^{n+1} - (K_{ion} + K_{ele} + W_B)^{n}}{\Delta t}$ may be written in three components:

\begin{align}
    \begin{split}
         &\sum_{i}\Delta X\left[-\frac{\gamma}{\gamma-1}\left(\nabla\cdot\left(\mathbf{u}_ep_e\right)\right)^{n+\frac{1}{2}}_i\right] \\
         + &\sum_{i}\Delta X\left[\mathbf{u}_i^{n+\frac{1}{2}}\cdot\left(\mathbf{J}\times\left(\nabla\times\mathbf{A}\right)\right)_i^{n+\frac{1}{2}} + \left(\mathbf{u}_i^{n+\frac{1}{2}}\times\left(\nabla\times\mathbf{A}\right)_i^{n+\frac{1}{2}}\right)\cdot\mathbf{J}_i^{n+\frac{1}{2}}\right] \\
         + &\sum_{i}\Delta X\left[\mathbf{u}_{e,i}^{n+\frac{1}{2}} - \mathbf{u}_{i}^{n+\frac{1}{2}} + \frac{1}{n_i^{n+\frac{1}{2}}e}\mathbf{J}^{n+\frac{1}{2}}_i\right]\cdot\left(\nabla p_e\right)_{i}^{n+\frac{1}{2}}.
    \end{split}
\end{align}

The first sums to zero assuming certain boundary conditions, the second uses the same vector identity as the continuous case to go to zero, and the third is zero by definition of $\mathbf{J}$. Thus, given proper boundary conditions, total energy is conserved.
A short aside on the Lagrange multipliers: throughout this section we have mentioned that discrete IBP on a finite domain may result in errors, offering Lagrange multipliers, discussed more thoroughly in Appendix \ref{appendix:lagrange-multipliers}, as a solution to this problem. It is important to emphasize that our method in itself is conserving, but with finite domains a conserving method may have spurious information added from the domain due to errors from discrete IBP. This would happen to any discretization scheme that relies on IBP, and does not reflect anything inherent to our method aside from this reliance. Additionally, we emphasize that applying the Lagrange multipliers to a non-conserving method, or to our conserving method with the standard truncation scheme (see Section \ref{section:tensor-trains}), will not aid in conserving mass, momentum, or energy, it only serves to correct the errors that come with discrete IBP.

\section{Tensor Trains} \label{section:tensor-trains}
\subsection{Introduction}

As discussed in Section \ref{section:introduction}, low-rank tensor representations have been enjoying something of a renaissance in recent years \cite{Kolda_Bader_2009}. In 2011 Oseledets developed a new method of decomposing a tensor known as a Tensor Train (TT) \cite{oseledets_2011}, avoiding the need to manage the Hierarchical Tucker's \cite{Hackbusch_Kuhn_2009, Grasedyck_2010} tree structure while not increasing memory footprint.\footnote{A na\"{i}ve implementation does give a slightly worse footprint, but Oseledets provides a way of recovering an identical footprint.} It is in this decomposition that this section is interested. We will go over some of the properties of TTs in Section \ref{subsection:tensor-trains-properties}, followed by a more detailed look into the method of truncation in Section \ref{subsection:tensor-train-truncation}. We will then explore a conservative truncation method and improve upon it in Section \ref{subsection:conserving-truncation}.
\subsection{Tensor Train Properties} \label{subsection:tensor-trains-properties}

As discussed in Section \ref{section:introduction}, there are a number of ways to store a low-rank representation of a tensor, one of the more recent developments being the Tensor Train decomposition \cite{oseledets_2011}, which effectively unfolds a $d$-dimensional tensor into a summation of the $r$ products of $d$ matrices\footnote{Here we are using a script letter to denote a tensor, a capital letter to denote a matrix, a boldfaced lowercase letter to denote a vector, and a nonboldfaced lowercase letter to denote a scalar. The indices for our examples are (row, depth, column), starting from the upper-front-left corner.}
\begin{equation}
    \mathcal{A}\left(i_1,i_2,...,i_d\right) = \sum_{\alpha_0,\alpha_1,...,\alpha_d}\prod_{\beta=1}^d\mathcal{G}^{(\beta)}\left(\alpha_{\beta-1}, i_\beta, \alpha_{\beta}\right) = \prod_{\beta=1}^d\mathcal{G}^{(\beta)}(i_\beta).
\end{equation}

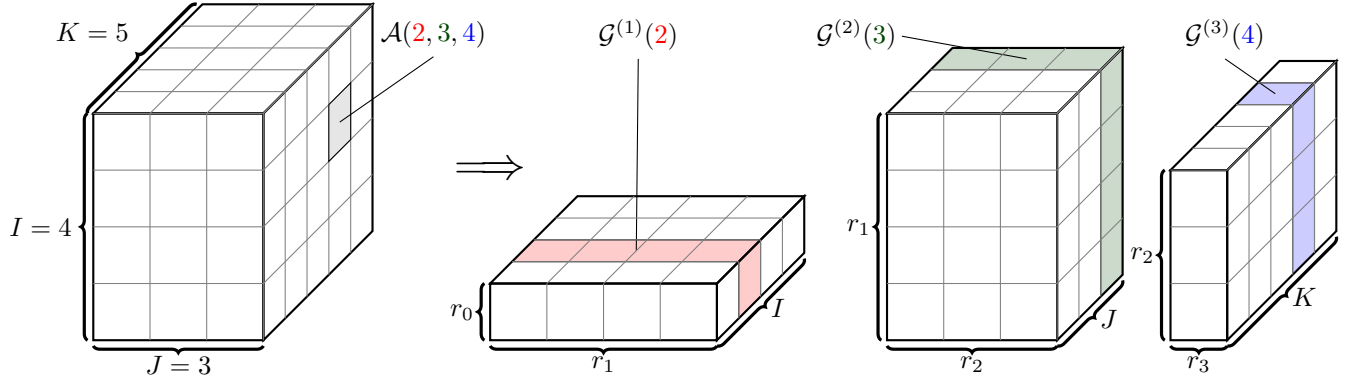
\begin{figure}
    \centering
    
    \begin{tikzpicture}[scale=.75]
        \def\o{(0,0,0)}
        \def\i{3}
        \def\j{4}
        \def\k{-5}

        \coordinate (O) at (0,0,0);
        \coordinate (I) at (\i,0,0);
        \coordinate (J) at (0,\j,0);
        \coordinate (K) at (0,0,\k);

        \def\kp1{\numexpr\k+1}
        % Draw outer cube
        \draw[thick] (O) -- (J) -- ($(J)+(I)$) -- (I) -- cycle;
        \draw[thick] (J) -- ($(J)+(K)$) -- ($(J)+(K)+(I)$) -- ($(J)+(I)$) -- cycle;
        \draw[thick] (I) -- ($(I)+(K)$) -- ($(J)+(K)+(I)$) -- ($(J)+(I)$) -- cycle;

        % Draw grid lines in y-direction and z-direction
        \foreach \x in {1,...,\i} {
            \draw[gray] (\x,0,0) -- ($(\x,0,0) + (J)$);
            \draw[gray] ($(\x,0,0) + (J)$) -- ($(\x,0,0) + (J) + (K)$);
        }
        
        % Draw grid lines in x-direction and z-direction
        \foreach \y in {1,...,\j} {
            \draw[gray] (0,\y,0) -- ($(0,\y,0) + (I)$);
            \draw[gray] ($(0,\y,0) + (I)$) -- ($(0,\y,0) + (I) + (K)$);
        }
        
        % Draw grid lines in x-direction and y-direction
        \foreach \z in {-1,...,\kp1} {
            \draw[gray] ($(0,0,\z) + (J)$) -- ($(0,0,\z) + (J) + (I)$);
            \draw[gray] ($(0,0,\z) + (I)$) -- ($(0,0,\z) + (J) + (I)$);
        }

        \draw[decorate, decoration={brace}, line width=1.2pt] (-.1,0,0) -- (-.1,4,0) node[midway, left=2pt] {$I=4$};
        \draw[decorate, decoration={brace}, line width=1.2pt] (3,-.1,0) -- (0,-.1,0) node[midway, below] {$J=3$};
        \draw[decorate, decoration={brace}, line width=1.3pt] (-.1,4,0) -- (-.1,4,-5) node[midway, left=1pt, yshift=10pt] {$K=5$};

        \filldraw[fill=gray!20, draw=black] (\i,2,-3) -- (\i,3,-3) -- (\i,3,-4) -- (\i,2,-4) -- cycle;

        \draw[-] (\i, 2.5, -3.5) -- (6,5, 0) node[above] {$\mathcal{A}(\textcolor{red}{2},\textcolor{DarkGreen}{3},\textcolor{blue}{4})$};

        \node[font=\LARGE] at (7,3) {$\implies$};

        \coordinate (O_new) at (\numexpr\i+4,0,0);

        \begin{scope}[shift={(O_new)}]

            \coordinate (O) at (0,0,0);
            \coordinate (I) at (\j,0,0); % intentional, i and j are kinda swapped here
            \coordinate (J) at (0,1,0);
            \coordinate (K) at (0,0,-4);

            \filldraw[fill=red!20, draw=black] (\j,0,-1) -- (\j,1,-1) -- (\j,1,-2) -- (\j,0,-2) -- cycle;
            \filldraw[fill=red!20, draw=black] (0,1,-1) -- (0,1,-2) -- (\j,1,-2) -- (\j,1,-1) -- cycle;

            \draw[decorate, decoration={brace}, line width=1.2pt] (4,-.1,-4) -- (4,-.1,0) node[midway, below, right, yshift=-2pt] {$I$};
            \draw[decorate, decoration={brace}, line width=1.2pt] (-.1,0,0) -- (-.1,1,0) node[midway, left] {$r_0$};
            \draw[decorate, decoration={brace}, line width=1.2pt] (4,-.1,0) -- (0,-.1,0) node[midway, below] {$r_1$};
    
            \draw[-] (2, 1, -1.5) -- (1.65, 4, -2.5) node[above] {$\mathcal{G}^{(1)}(\textcolor{red}{2})$};

            % Draw outer cube
            \draw[thick] (O) -- (J) -- ($(J)+(I)$) -- (I) -- cycle;
            \draw[thick] (J) -- ($(J)+(K)$) -- ($(J)+(K)+(I)$) -- ($(J)+(I)$) -- cycle;
            \draw[thick] (I) -- ($(I)+(K)$) -- ($(J)+(K)+(I)$) -- ($(J)+(I)$) -- cycle;

            % Draw grid lines in y-direction and z-direction
            \foreach \x in {1,...,\i} {
                \draw[gray] (\x,0,0) -- ($(\x,0,0) + (J)$);
                \draw[gray] ($(\x,0,0) + (J)$) -- ($(\x,0,0) + (J) + (K)$);
            }
            
            % % Draw grid lines in x-direction and y-direction
            \foreach \z in {-1,...,-3} {
                \draw[gray] ($(0,0,\z) + (J)$) -- ($(0,0,\z) + (J) + (I)$);
                \draw[gray] ($(0,0,\z) + (I)$) -- ($(0,0,\z) + (J) + (I)$);
            }

        \end{scope}

        \coordinate (O_new) at (\numexpr\i+11,0,0);

        \begin{scope}[shift={(O_new)}]

            \coordinate (O) at (0,0,0);
            \coordinate (I) at (\i,0,0);
            \coordinate (J) at (0,\j,0);
            \coordinate (K) at (0,0,-3);

            \filldraw[fill=DarkGreen!20, draw=black] (\i,0,-2) -- (\i,4,-2) -- (\i,4,-3) -- (\i,0,-3) -- cycle;
            \filldraw[fill=DarkGreen!20, draw=black] (0,4,-2) -- (0,4,-3) -- (\i,4,-3) -- (\i,4,-2) -- cycle;

            % \draw[-] (1.5, 4, -2.5) -- (-1.5, 4, -2.5) node[above] {$\mathcal{G}^{(2)}(\textcolor{DarkGreen}{3})$};
            \draw[-] (1.5, 4, -2.5) -- (-0.7, 4.4, -2.5);
            \node[above] at (-1.5, 4, -2.5) {$\mathcal{G}^{(2)}(\textcolor{DarkGreen}{3})$};

            \draw[decorate, decoration={brace}, line width=1.2pt] (3,-.1,-3) -- (3,-.1,0) node[midway, below, right, yshift=-2pt] {$J$};
            \draw[decorate, decoration={brace}, line width=1.2pt] (-.1,0,0) -- (-.1,4,0) node[midway, left] {$r_1$};
            \draw[decorate, decoration={brace}, line width=1.2pt] (3,-.1,0) -- (0,-.1,0) node[midway, below] {$r_2$};

            % Draw outer cube
            \draw[thick] (O) -- (J) -- ($(J)+(I)$) -- (I) -- cycle;
            \draw[thick] (J) -- ($(J)+(K)$) -- ($(J)+(K)+(I)$) -- ($(J)+(I)$) -- cycle;
            \draw[thick] (I) -- ($(I)+(K)$) -- ($(J)+(K)+(I)$) -- ($(J)+(I)$) -- cycle;

            % Draw grid lines in y-direction and z-direction
            \foreach \x in {1,...,\i} {
                \draw[gray] (\x,0,0) -- ($(\x,0,0) + (J)$);
                \draw[gray] ($(\x,0,0) + (J)$) -- ($(\x,0,0) + (J) + (K)$);
            }

            % Draw grid lines in x-direction and z-direction
            \foreach \y in {1,...,\j} {
                \draw[gray] (0,\y,0) -- ($(0,\y,0) + (I)$);
                \draw[gray] ($(0,\y,0) + (I)$) -- ($(0,\y,0) + (I) + (K)$);
            }
            
            % Draw grid lines in x-direction and y-direction
            \foreach \z in {-1,...,-2} {
                \draw[gray] ($(0,0,\z) + (J)$) -- ($(0,0,\z) + (J) + (I)$);
                \draw[gray] ($(0,0,\z) + (I)$) -- ($(0,0,\z) + (J) + (I)$);
            }

        \end{scope}

        \coordinate (O_new) at (\numexpr\i+16,0,0);

        \begin{scope}[shift={(O_new)}]

            \coordinate (O) at (0,0,0);
            \coordinate (I) at (1,0,0);
            \coordinate (J) at (0,\i,0); % intentional, i and j are swapped
            \coordinate (K) at (0,0,\k);

            \filldraw[fill=blue!20, draw=black] (1,0,-3) -- (1,3,-3) -- (1,3,-4) -- (1,0,-4) -- cycle;
            \filldraw[fill=blue!20, draw=black] (0,3,-3) -- (0,3,-4) -- (1,3,-4) -- (1,3,-3) -- cycle;

            \draw[-] (0.5, 3, -3.5) -- (0, 4, -2.5) node[above] {$\mathcal{G}^{(3)}(\textcolor{blue}{4})$};

            \draw[decorate, decoration={brace}, line width=1.2pt] (1,-.1,-5) -- (1,-.1,0) node[midway, below, right, yshift=-2pt] {$K$};
            \draw[decorate, decoration={brace}, line width=1.2pt] (-.1,0,0) -- (-.1,3,0) node[midway, left] {$r_2$};
            \draw[decorate, decoration={brace}, line width=1.2pt] (1,-.1,0) -- (0,-.1,0) node[midway, below] {$r_3$};

            % Draw outer cube
            \draw[thick] (O) -- (J) -- ($(J)+(I)$) -- (I) -- cycle;
            \draw[thick] (J) -- ($(J)+(K)$) -- ($(J)+(K)+(I)$) -- ($(J)+(I)$) -- cycle;
            \draw[thick] (I) -- ($(I)+(K)$) -- ($(J)+(K)+(I)$) -- ($(J)+(I)$) -- cycle;

            % Draw grid lines in x-direction and z-direction
            \foreach \y in {1,...,\i} {
                \draw[gray] (0,\y,0) -- ($(0,\y,0) + (I)$);
                \draw[gray] ($(0,\y,0) + (I)$) -- ($(0,\y,0) + (I) + (K)$);
            }
            
            % Draw grid lines in x-direction and y-direction
            \foreach \z in {-1,...,\kp1} {
                \draw[gray] ($(0,0,\z) + (J)$) -- ($(0,0,\z) + (J) + (I)$);
                \draw[gray] ($(0,0,\z) + (I)$) -- ($(0,0,\z) + (J) + (I)$);
            }

        \end{scope}
    \end{tikzpicture}
    \caption{The full-rank TT decomposition of an $\mathbb{R}^{4\times3\times5}$ tensor. Figure inspired by \cite{xu2023tensor}. Observe that $\mathcal{G}^{(1)}(\textcolor{red}{2}) \in \mathbb{R}^{r_0\times r_1}$, $\mathcal{G}^{(2)}(\textcolor{DarkGreen}{3}) \in \mathbb{R}^{r_1 \times r_2}$, and $\mathcal{G}^{(3)}(\textcolor{blue}{4}) \in \mathbb{R}^{r_2\times r_3}$, making the product $\mathcal{G}^{(1)}(\textcolor{red}{2})\mathcal{G}^{(2)}(\textcolor{DarkGreen}{3})\mathcal{G}^{(3)}(\textcolor{blue}{4})$ an $r_0\times r_3$ matrix. If we assert $r_0=r_3=1$, then this becomes a scalar, which is exactly what we want, as this is equal to $\mathcal{A}\left(\textcolor{red}{2},\textcolor{DarkGreen}{3},\textcolor{blue}{4}\right) \in \mathbb{R}$.}
    \label{fig:3d-tensor}
\end{figure}

We write the second term as an indexed summation, made more pithy by the third term, in which $\mathcal{G}^{(\sigma)}(i_\sigma)\mathcal{G}^{(\tau)}(i_\tau)$ is a matrix product. Given $\mathcal{A}\left(i_1,i_2,...,i_d\right)$ is a scalar, we require $\mathcal{G}^{(1)}$ to be a $1 \times J_1 \times r_1$ tensor and $\mathcal{G}^{(d)}$ to be a $r_{d-1}\times J_d \times 1$ tensor. In other words, the slices along the second dimension result in row and column matrices, forcing the end product to be a scalar. See Figure \ref{fig:3d-tensor}.

\begin{table}[ht]
    \centering
    \begin{tabular}{lcc}
    \hline
    Operation & Resulting TT Rank & Computational Cost \\
    \hline
    Na\"ive inner product $\langle \mathcal{A},\mathcal{B}\rangle$
    &
    --
    &
    $\mathcal{O}\!\left(d\,n\,r_\mathcal{A}^2 r_\mathcal{B}^2\right)$
    \\[0.5em]

    Improved inner product $\langle \mathcal{A},\mathcal{B}\rangle$
    &
    --
    &
    $\mathcal{O}\!\left(d\,n\,r_\mathcal{A} r_\mathcal{B}\left(r_\mathcal{A} + r_\mathcal{B}\right)\right)$
    \\[0.5em]

    Addition $\mathcal{A}+\mathcal{B}$
    &
    $r_\mathcal{A}+r_\mathcal{B}$
    &
    $\mathcal{O}\!\left(d\,n\,(r_\mathcal{A}+r_\mathcal{B})^2\right)$
    \\[0.5em]

    Hadamard product $\mathcal{A}\star \mathcal{B}$
    &
    $r_\mathcal{A} r_\mathcal{B}$
    &
    $\mathcal{O}\!\left(d\,n\,r_\mathcal{A}^2 r_\mathcal{B}^2\right)$
    \\[0.5em]

    Scalar/Tensor addition $s+\mathcal{A}$
    &
    $r_\mathcal{A} + 1$
    &
    $\mathcal{O}\!\left(d\,n\,(r_\mathcal{A}+1)^2\right)$
    \\[0.5em]

    Scalar/Tensor product $s\mathcal{A}$
    &
    $r_\mathcal{A}$
    &
    $\mathcal{O}\!\left(d\,n\,r_\mathcal{A}^2\right)$
    \\[0.5em]

    TT truncation of $\mathcal{A}$ to rank $r_{\text{post}}$
    &
    $r_{\text{post}}$
    &
    $\mathcal{O}\!\left(d\,n\,r_\mathcal{A}^3\right)$
    \\
    \hline
    \end{tabular}
    \caption{Computational complexity and rank growth of common TT operations for TT objects $\mathcal{A}$ and $\mathcal{B}$ and scalar $s$. Assumes uniform mode size $n$, dimension $d$, and interior ranks $r_\mathcal{A}$ and $r_\mathcal{B}$ for $\mathcal{A}$ and $\mathcal{B}$, respectively \cite{oseledets_2011}.}
    \label{tab:tensor-operations-computational-complexity}
\end{table}

Simple operations increase the rank of this object. For example, the addition of two tensors $\mathcal{C} = \mathcal{A}+\mathcal{B}$ is performed
\begin{equation*}
    \mathcal{G}_{\mathcal{C}}^{(1)}(i_1) = \begin{pmatrix}
        \mathcal{G}_{\mathcal{A}}^{(1)}(i_1) & \mathcal{G}_{\mathcal{B}}^{(1)}(i_1)
    \end{pmatrix},
    \quad
    \mathcal{G}_{\mathcal{C}}^{(k)}(i_k) = \begin{pmatrix}
        \mathcal{G}_{\mathcal{A}}^{(k)}(i_k) & 0 \\ 0 & \mathcal{G}_{\mathcal{B}}^{(k)}(i_k)
    \end{pmatrix} 2 \leq k\leq d-1,
    \quad
    \mathcal{G}_{\mathcal{C}}^{(d)}(i_d) = \begin{pmatrix}
        \mathcal{G}_{\mathcal{A}}^{(d)}(i_d) \\ \mathcal{G}_{\mathcal{B}}^{(d)}(i_d)
    \end{pmatrix}.
\end{equation*}

In other words, the new tensor's rank is increased to $r_\mathcal{C} = r_\mathcal{A} + r_\mathcal{B}$. The Hadamard (elementwise) product, $\mathcal{C} = \mathcal{A}\star\mathcal{B}$, affects $\mathcal{C}$'s cores as follows
\begin{equation*}
    \mathcal{G}_{\mathcal{C}}^{(k)}\left(i_k\right) = \mathcal{G}_\mathcal{A}^{(k)}(i_k) \otimes \mathcal{G}_\mathcal{B}^{(k)}(i_k).
\end{equation*}

Here $\otimes$ is the Kronecker product of matrices.\footnote{Ref. \cite{Henderson_etal_1983} argues that the Kronecker product is more properly attributed to lesser-known mathematician Johann Georg Zehfuss.} Accordingly, the new tensor's rank is $r_\mathcal{C} = r_\mathcal{A}r_\mathcal{B}$. 

Oseledets goes into these, and other, operations in further detail (Table \ref{tab:tensor-operations-computational-complexity} gives the operations relevant to our project). In short, even a few relatively minor operations result in an unacceptably large rank, potentially leaving us worse off than we would have been had we simply kept to the full-rank tensor. However, Oseledets does not leave us without a fix for this.

\subsection{Tensor Train Truncation} \label{subsection:tensor-train-truncation}

Just as a matrix $A$ may be decomposed into three matrices $U\Sigma V^T$ via the singular value decomposition (SVD), so our $d$-dimensional tensor $\mathcal{A}$ is decomposed into a product of $d$ $3$-dimensional core tensors $\prod_{k=1}^d\mathcal{G}^{(k)}$ ($2$ dimensions for the first and last core tensor) by the TT decomposition. And just as we may truncate the SVD, so also we may truncate the TT decomposition. In fact, the methodology of truncating the TT is predicated on use of the SVD. Following \cite{oseledets_2011}, given a tensor $\mathcal{A}$, we can decompose this into a \enquote{train} $\tilde{\mathcal{A}}$ of three-dimensional tensors \enquote{bounded} by two-dimensional tensors (Figure \ref{fig:3d-tensor}), truncating as we go. Should we already have the tensor in the form $\tilde{\mathcal{A}}$ and apply some operation to it, we can truncate again using \enquote{TT-rounding}, which is more efficient than applying a decomposition repeatedly.

Should $r$ be considerably lower than $N$, we find this provides a satisfying truncation to $\mathcal{A}$, reducing the space requirements from $\mathcal{O}\left(N^d\right)$ to $\mathcal{O}\left(2Nr + (d - 2)Nr^2\right)$, and with some small effort, reduced to $\mathcal{O}\left(dNr + (d-2)r^3\right)$.

It is provable that $\lVert \mathcal{A} - \tilde{\mathcal{A}}\rVert_F \leq \sqrt{\sum_{k=1}^{d-1}\epsilon_{k}^2}$, where $\epsilon_k$ is the truncation error in the $k$th unfolding of $\mathcal{A}$ \cite{oseledets_2011}. This method is relatively fast and intuitive, but does not discriminate what information it cuts. If we wish to maintain conservation of our moments, we cannot use this blindly. It is to addressing this issue we now turn.

% \subsubsection{A Conserving Truncation Method} \label{subsection:gq-conserving-truncation}
\subsection{A Rank-2 Conserving Truncation Method} \label{subsection:conserving-truncation}

As mentioned above, the standard TT-SVD truncation makes no distinction as to the information it is truncating, it only cares about the size of the singular values. Guo and Qiu explored how to conserve the first three moments of a distribution function while still keeping the rank of the tensor object a reasonable size \cite{guo_qiu_2024_1, guo_qiu_2024_2, heenatigala_guo_2026}. While they truncate their tensor objects using HT decomposition \cite{Hackbusch_Kuhn_2009, Grasedyck_2010}, the method they use to preserve the moment information is perfectly relevant to our TT decomposition. The truncation method they develop requires two projection steps, each of which adds $7$ rank to the tensor object for a total of $14$. Here we will briefly discuss their work and how this may be reduced to an addition of $2$ rank.

Given a distribution function $f$\footnote{We reduce to a single species without loss of generality.} defined on the space and velocity domain $\Omega_{\mathbf{x}} \times \Omega_{\mathbf{v}}$, we find the mass, momentum, and (kinetic) energy moments at space $\mathbf{x}$ and time $t$ taking the moments of the distribution function \eqref{eq:mass-moment-continuous-s}-\eqref{eq:KE-moment-continuous-s}. We discretize and find we have a distribution function $\mathbf{f}$ and a discretized phase space $\mathbf{x} \otimes \mathbf{y} \otimes \mathbf{z} \otimes \mathbf{v}_x \otimes \mathbf{v}_y \otimes \mathbf{v}_z$ (see Section \ref{subsection:discrete-hybrid-model} for more details); we utilize the standard inner product representing the integral operator in discrete space \eqref{eq:discrete-integral}.

We see \eqref{eq:mass-moment-continuous-s}-\eqref{eq:KE-moment-continuous-s} may be written as:
\begin{align}
    \rho &= \langle\mathbf{f}, \mathbf{1}_{v_x} \otimes \mathbf{1}_{v_y} \otimes \mathbf{1}_{v_z}\rangle \equiv \langle\mathbf{f}, \mathbf{V}_1\rangle, \\
    J_1 &= \langle\mathbf{f}, \mathbf{v}_x \otimes \mathbf{1}_{v_y} \otimes \mathbf{1}_{v_z}\rangle \equiv \langle\mathbf{f}, \mathbf{V}_2\rangle, \\
    J_2 &= \langle\mathbf{f}, \mathbf{1}_{v_x} \otimes \mathbf{v}_y \otimes \mathbf{1}_{v_z}\rangle \equiv \langle\mathbf{f}, \mathbf{V}_3\rangle, \\
    J_3 &= \langle\mathbf{f}, \mathbf{1}_{v_x} \otimes \mathbf{1}_{v_y} \otimes \mathbf{v}_z\rangle \equiv \langle\mathbf{f}, \mathbf{V}_4\rangle, \\
    \begin{split}
        \kappa &= \frac{1}{2}\langle\mathbf{f}, \mathbf{v}_x^2 \otimes \mathbf{1}_{v_y} \otimes \mathbf{1}_{v_z}\rangle + \frac{1}{2}\langle\mathbf{f}, \mathbf{1}_{v_x} \otimes \mathbf{v}_y^2 \otimes \mathbf{1}_{v_z}\rangle + \frac{1}{2}\langle\mathbf{f}, \mathbf{1}_{v_x} \otimes \mathbf{1}_{v_y} \otimes \mathbf{v}_z^2\rangle \\
         &\equiv \frac{1}{2}\langle\mathbf{f}, \mathbf{V}_{\kappa,1}\rangle + \frac{1}{2}\langle\mathbf{f}, \mathbf{V}_{\kappa,2}\rangle + \frac{1}{2}\langle\mathbf{f}, \mathbf{V}_{\kappa,3}\rangle \\
         &\equiv \frac{1}{2}\langle\mathbf{f}, \mathbf{V}_{\kappa}\rangle.
    \end{split}
\end{align}

Here we have defined $\mathbf{v}_\eta^2 \coloneqq \mathbf{v}_\eta \star \mathbf{v}_\eta$ for $\eta \in \{x,y,z\}$, recalling $\star$ is the Hadamard product. The $\mathbf{V}_\kappa$ notation is intentional, we will define $\mathbf{V}_5$ and employ it now. 
\begin{align}
    \begin{split}
        \mathbf{V}_5 &\coloneqq \mathbf{V}_{5,1} + \mathbf{V}_{5,2} + \mathbf{V}_{5,3} \\
        &\coloneqq \left(\mathbf{v}_x^2 - c_x\mathbf{1}_{v_x}\right) \otimes \mathbf{1}_{v_y} \otimes \mathbf{1}_{v_z} + \mathbf{1}_{v_x} \otimes (\mathbf{v}_y^2 - c_y\mathbf{1}_{v_y}) \otimes \mathbf{1}_{v_z} + \mathbf{1}_{v_x} \otimes \mathbf{1}_{v_y} \otimes \left(\mathbf{v}_z^2 - c_z\mathbf{1}_{v_z}\right), \\
        c_\eta &\coloneqq \frac{\langle\bar{\mathbf{1}}, \bar{\mathbf{v}}_\eta^2\rangle}{\langle\bar{\mathbf{1}},\bar{\mathbf{1}}\rangle}.
    \end{split}
\end{align}

Here a bar denotes the embedding of a one-dimensional factor into the full velocity space, formed by tensoring with ones in the remaining directions, e.g. $\bar{\mathbf{v}}_{x}^2 \coloneqq \mathbf{v}_x^2\otimes\mathbf{1}_{v_y}\otimes\mathbf{1}_{v_z}$. Now, the reason we have defined $\mathbf{V}_5$ in this way is so that the following subspace is orthogonal:
\begin{align}
    \mathcal{V} \coloneqq \{\mathbf{V}_1, \mathbf{V}_2, \mathbf{V}_3, \mathbf{V}_4, \mathbf{V}_5\}.
\end{align}

We acquire the portion of $\mathbf{f}$ containing all the information of the first three moments by projecting it onto $\mathcal{V}$. These basis vectors have nonzero values along the boundaries, which causes issues with a na\"{i}ve projection. As such, we use a weighted projection, typically with weight $\mathbf{w}(v_x,v_y,v_z) = \exp\left(-\frac{v_x^2 + v_y^2 + v_z^2}{2v_{th}^2}\right)$ (see \cite{zheng_sands_qiu_2026} for a more flexible weighting scheme), which forces the subspace vectors to zero along the boundaries. We define the weighted inner product and induced weighted norm
\begin{align}
    &\langle\mathbf{a},\mathbf{b}\rangle_\mathbf{w} \coloneqq \Delta v_x\Delta v_y\Delta v_z\sum_{i,j,k}a_{i,j,k}b_{i,j,k}w_{i,j,k}, \\
    &\left\lVert\mathbf{a}\right\rVert_\mathbf{w} \coloneqq \sqrt{\langle\mathbf{a},\mathbf{a}\rangle_\mathbf{w}}.
\end{align}

With this weighted inner product and norm defined, we redefine the constants $c_\eta$ to use them, $c_\eta \coloneqq \langle\bar{\mathbf{1}}, \bar{\mathbf{v}}_\eta^2\rangle_\mathbf{w} \big/ \left\lVert\bar{\mathbf{1}}\right\rVert^2_\mathbf{w}$. This ensures $\mathbf{V}_5$ is orthogonal to $\mathbf{V}_1$ under $\langle\cdot,\cdot\rangle_\mathbf{w}$, so that the energy correction does not perturb the density.

There is a shortcut worth noting. First, if we were to scale $\mathbf{f} \rightarrow \frac{1}{\mathbf{w}}\star\mathbf{f} \equiv \hat{\mathbf{f}}$, there is no difference projecting this vector rather than the original and then simply rescaling afterwards. But, given we are projecting using the same weight vector $\mathbf{w}$, the dot product $\langle\cdot,\cdot\rangle_\mathbf{w}$ will \enquote{cancel} out with the scale factor to form a standard, unweighted, dot product, i.e.
\begin{equation}
    \left\langle\frac{1}{\mathbf{w}}\star\mathbf{f},\mathbf{M}\right\rangle_\mathbf{w} = \Delta v_x\Delta v_y\Delta v_z\sum_{i,j,k}\frac{1}{w_{i,j,k}}f_{i,j,k}M_{i,j,k}w_{i,j,k} = \Delta v_x\Delta v_y\Delta v_z\sum_{i,j,k}f_{i,j,k}M_{i,j,k} = \left\langle \mathbf{f},\mathbf{M}\right\rangle.
\end{equation}

The methodology of Guo and Qiu constructs a projector $\mathcal{P}_w$ using these weighted inner products, projects the moment information to keep it isolated $\mathbf{f}_1 = \mathcal{P}_w\left[\mathbf{f}\right]$ with the remainder being contained in the orthogonal complement $\mathbf{f}_2 = \left(\mathcal{I} - \mathcal{P}_w\right)\left[\mathbf{f}\right]$. It then truncates the orthogonal complement $\tilde{\mathbf{f}}_2 = \mathcal{T}\left[\mathbf{f}_2\right]$, and then projects any spurious moment information the truncation may have injected.\footnote{The reason this is necessary is that truncating the moment-less orthogonal complement of $\mathbf{f}_2$ may inject spurious moment information, knocking it out of the $\mathcal{V}^C$ space. An example may elucidate this: the vector $\mathbf{v} = (1,1,0) = \left(\frac{1}{2},\frac{1}{2},+1\right) + \left(\frac{1}{2},\frac{1}{2},-1\right)$ \enquote{lives} on $\mathbb{R}^2$ (strictly speaking it is the image of the canonical embedding $\iota: \mathbb{R}^2 \hookrightarrow \mathbb{R}^3$ applied to $(1,1)$). But should we cut the second vector, resulting in $\tilde{\mathbf{v}} = \left(\frac{1}{2},\frac{1}{2},1\right)$, we see we have knocked the vector off of the $\mathbb{R}^2$ space and into $\mathbb{R}^3$.} This results in adding $14$ rank, $7$ from each projection step. Instead we work in reverse, truncating and correcting whatever moment deviation has occurred. This is similar to the approach suggested in \cite{Kormann_2015}, which preserved mass and momentum by correcting the recently truncated tensor with moment information from the tensor at $t=0$. We have our distribution function $\mathbf{f}$, and compute the known $\rho$, $\mathbf{J}$, and $\kappa$ values. Now, we can truncate $\mathbf{f}_t = \mathcal{T}\left[\mathbf{f}\right]$, and compute the truncated moments $\rho_t$, $\mathbf{J}_t$, and $\kappa_t$. We compute the differences $\rho_d = \rho - \rho_t$, $\mathbf{J}_d = \mathbf{J} - \mathbf{J}_t$, and $\kappa_d = \kappa - \kappa_t$. Knowing these differences, we can compute a correction term.
\begin{equation}
    \mathbf{f}_c = \frac{\rho_d}{\lVert\mathbf{V}_1\rVert_\mathbf{w}^2}\left(\mathbf{V}_1 \star \mathbf{w}\right) 
      + \frac{J_{d,1}}{\lVert\mathbf{V}_2\rVert_\mathbf{w}^2}\left(\mathbf{V}_2 \star \mathbf{w}\right) + \frac{J_{d,2}}{\lVert\mathbf{V}_3\rVert_\mathbf{w}^2}\left(\mathbf{V}_3 \star \mathbf{w}\right) + \frac{J_{d,3}}{\lVert\mathbf{V}_4\rVert_\mathbf{w}^2}\left(\mathbf{V}_4 \star \mathbf{w}\right)
      + \frac{2\kappa_d - (c_x+c_y+c_z)\rho_d}{\lVert\mathbf{V}_5\rVert_\mathbf{w}^2}\left(\mathbf{V}_5 \star \mathbf{w}\right)
\end{equation}

We simply add this information back into the truncated distribution function, $\mathbf{f}_{conserved} = \mathbf{f}_t + \mathbf{f}_c$. This leaves us with a minimum rank of $\tilde{r} + 7$ rather than $\tilde{r} + 14$, as it only requires one projection step rather than two. We can further improve this. Noting how many of these tensors share values across dimensions, we rewrite a few constants and then rearrange:
\begin{align} 
    &\left(\left(\left(\tilde{K}_1\mathbf{1}_{v_x} + K_2\mathbf{v}_x + K_5\mathbf{v}_x^2\right)\otimes\mathbf{1}_{v_y}\otimes\mathbf{1}_{v_z}\right) + \left(\mathbf{1}_{v_x}\otimes\left(K_3\mathbf{v}_y + K_5\mathbf{v}_y^2\right)\otimes\mathbf{1}_{v_z}\right) + \left(\mathbf{1}_{v_x}\otimes\mathbf{1}_{v_y}\otimes\left(K_4\mathbf{v}_z + K_5\mathbf{v}_z^2\right)\right)\right)\star\mathbf{w}, \\
    &K_1 \coloneqq \frac{\rho_d}{\lVert\mathbf{V}_1\rVert_\mathbf{w}^2}, K_2 \coloneqq \frac{J_{d,1}}{\lVert\mathbf{V}_2\rVert_\mathbf{w}^2}, K_3 \coloneqq \frac{J_{d,2}}{\lVert\mathbf{V}_3\rVert_\mathbf{w}^2}, K_4 \coloneqq \frac{J_{d,3}}{\lVert\mathbf{V}_4\rVert_\mathbf{w}^2}, K_5 \coloneqq \frac{2\kappa_d - (c_x+c_y+c_z)\rho_d}{\lVert\mathbf{V}_5\rVert_\mathbf{w}^2},  \label{eq:K_variables} \\
    &\tilde{K}_1 \coloneqq K_1 - (c_x+c_y+c_z)K_5. \label{eq:K_1_tilde_variable}
\end{align}

So instead of adding $7$ ranks, we now only add $3$. Our situation is slightly different from this, however. Recalling we are not interested in the canonical decomposition, but rather the TT decomposition, we can rewrite this as follows:

\begin{equation} \label{eq:correction_tensor_rank_2}
    \mathbf{f}_c = 
    \left[
    \begin{pmatrix}
        \mathbf{1}_{v_x} & K_2\mathbf{v}_x + K_5\mathbf{v}_x^2
    \end{pmatrix}
    \begin{pmatrix}
        \mathbf{1}_{v_y} & K_3\mathbf{v}_y + K_5\mathbf{v}_y^2 \\
        \mathbf{0} & \mathbf{1}_{v_y}
    \end{pmatrix} \\
    \begin{pmatrix}
       \tilde{K}_1\mathbf{1}_{v_z} + K_4\mathbf{v}_z + K_5\mathbf{v}_z^2 \\
        \mathbf{1}_{v_z}
    \end{pmatrix}
    \right]\star\mathbf{w}.
\end{equation}

These products are connected with a Kronecker product of matrices.\footnote{e.g. $(\mathbf{a} \quad \mathbf{b})\begin{pmatrix}\mathbf{c}\\\mathbf{d}\end{pmatrix} = \mathbf{a}\otimes\mathbf{c} + \mathbf{b}\otimes\mathbf{d}$} This is a TT object with rank $[1, 2, 2, 1]$.

\section{Implementation} \label{section:implementation}

Section \ref{section:hybrid-model} gave us the foundation of the model we wish to simulate as well as its discretization scheme. Section \ref{section:tensor-trains} gave us the ability to control the memory footprint. Now we turn to discuss some details of the implementation of this model.
\subsection{The Conservative Low-Rank Algorithm} \label{section:conservative-lowrank-alg}

Algorithm \ref{alg:hybrid-method} is the overall hybrid method, which uses a fixed-point iteration to update the fluid electrons and vector potential, enslaving the kinetic Vlasov solver to update the ions in step. The implementation is similar to that of \cite{stanier2019fully}. The details for the Vlasov solver may be found in Algorithm \ref{alg:CN_Vlasov}. This solver utilizes the conservative truncation scheme discussed in Section \ref{subsection:conserving-truncation}.

The total solution vector for our electromagnetic field formulation is $\hat{\mathcal{S}}^{n+1} \coloneqq \left(f_s^{n+1}, \mathbf{A}^{n+1}, p_{e}^{n+1}\right)$, but rather than solve for this, we imitate \cite{stanier2019fully}, implementing the so-called particle-enslavement method of \cite{Chen_etal_2011, Chen_chacon_2014}. Though they use this in the context of PIC, the principle remains the same: we update the kinetic component $f_s$ as the fields $\mathbf{A}, p_e$ are iterated upon, rather than iterating on $f_s$ directly, which is prohibitively large. This may be thought of as rewriting the residual $\mathbf{F}\left(\mathbf{X}_1,\mathbf{X}_2\right)$ into a new residual, $\mathbf{G}\left(\mathbf{X}_1\right)$, with $\mathbf{X}_2$ being a function of $\mathbf{X}_1$. In our case we have $\mathbf{X}_1 = (\mathbf{A}, p_e)$ and $\mathbf{X}_2 = f_s$, recovering the residual $\mathbf{H}(\mathcal{S})$, with our new solution vector being $\mathcal{S} \coloneqq (\mathbf{A}, p_e)$. This sidesteps the need to iterate on the distribution function, updating $f_s$ to be consistent with the fields and electron pressure instead.

At timestep $n$, iteration $k+1$, we compute
\begin{align}
    \mathbf{H}\left(\mathcal{S}^{n+1,(k)}\right) &= \begin{pmatrix}
        \dfrac{\mathbf{A}^{n+1,(k)} - \mathbf{A}^n}{\Delta t} + \mathbf{E}^{n+\theta,(k)} \\[2ex]
        \dfrac{p_e^{n+1,(k)} - p_e^n}{\Delta t} + p_{e,adv}^{n+\theta,(k)}
    \end{pmatrix}, \\
    p_{e,adv}^{n+\theta,(k)} &\coloneqq \gamma \nabla\cdot\left(p_e^{n+\theta,(k)}\,\mathbf{u}^{n+\theta,(k+1)}_e\right) - (\gamma-1)\,\mathbf{u}^{n+\theta,(k+1)}_e\cdot\nabla p_e^{n+\theta,(k)}. \label{eq:p_e_adv}
\end{align}

The superscript $n+\theta$ on the electric field $\mathbf{E}$ and the pressure advection $p_{e,adv}$ denotes that these nonlinear terms are each evaluated once, at the single blended state
\begin{equation}
    \mathcal{S}^{n+\theta,(k)} \coloneqq (1-\theta)\mathcal{S}^{n+1,(k)} + \theta\mathcal{S}^{n}.
\end{equation}

This differs from a $\theta$-weighted average of two separate evaluations at time levels $n$ and $n+1$, which for a nonlinear $\mathbf{E}$ or $p_{e,adv}$ would not coincide with a single evaluation at the blended state. Setting $\theta = \frac{1}{2}$ gives us implicit midpoint, which is needed for the conservation properties described above.

If we were only doing fixed-point iteration, we then assign this to the new solution vector.
\begin{equation}
    \mathcal{S}^{n+1,(k+1)} = \mathcal{S}^{n+1,(k)} - \Delta t\mathbf{H}\left(\mathcal{S}^{n+1,(k)}\right) \equiv \delta\mathcal{S}^{n+1,(k)}.
\end{equation}

We form our tolerance $\epsilon$ out of the absolute tolerance $\epsilon_a$ and the relative tolerance $\epsilon_r$, iterating until our new state is accepted:
\begin{equation}
    \left\lVert\mathbf{H}\left(\mathcal{S}^{n+1,(k+1)}\right)\right\rVert_2 < \epsilon \equiv \epsilon_a + \epsilon_r\left\lVert\mathbf{H}\left(\mathcal{S}^{n}\right)\right\rVert_2.
\end{equation}

While this technically works on its own, we implement Anderson Acceleration \cite{Anderson_1965, walker_ni_2011}. This is an extension of fixed-point iteration, taking an affine combination of the $m+1$ most recent fixed-point updates including the current (we set $m=5$),
\begin{equation}
    \mathcal{S}^{n+1,(k+1)} = \sum_{i=k-m}^k\alpha_{i-(k-m)}\delta\mathcal{S}^{n+1,(i)}.
\end{equation}

We add the constraint $\sum_{j}\alpha_j = 1$. $\vec{\alpha}^{(k)}$ is computed via least squares,
\begin{equation}
    \vec{\alpha}^{(k)} = \arg\min_{\sum_j\alpha_j=1}\left\lVert \sum_{i=k-m}^k\alpha_{i-(k-m)}\mathbf{H}\left(\mathcal{S}^{n+1,(i)}\right) \right\rVert_2
\end{equation}

The nonlinear iteration algorithm may be found in Algorithm \ref{alg:hybrid-method}, the implicit midpoint Vlasov solve in Algorithm \ref{alg:CN_Vlasov}. Both algorithms utilize the bracket notation $\langle\cdot,\cdot\rangle$ defined in \eqref{eq:discrete-integral}, denoting a discrete integral which takes the form of a dot product multiplied by some difference $\Delta$. Recall that we have set $\eta = 0$ for our purposes, meaning $\mathbf{E} = \mathbf{E}^*$. In Algorithm \ref{alg:hybrid-method} we nonetheless preserve the distinction between $\mathbf{E}$ and $\mathbf{E}^*$, passing $\mathbf{E}^*$ to Algorithm \ref{alg:CN_Vlasov}. Were $\eta \neq 0$, we would subtract the $\eta\mathbf{J}$ term from $\mathbf{E}$ to recover $\mathbf{E}^*$ before passing it.

\begin{algorithm}
\caption{Outline of the Hybrid Kinetic-Ion/Fluid-Electron Method}\label{alg:hybrid-method}
    \begin{algorithmic}

        \State \textbf{Given}: Collection of $N_x\times N_y\times N_z$ $d$-dimensional tensors $f_{ijk}(\mathbf{v}) \rightarrow f$, $N_x\times N_y \times N_z$ arrays $\left(\mathbf{A}_{ijk}, p_{e,ijk}\right) \rightarrow \left(\mathbf{A}, p\right)$, absolute and relative nonlinear tolerances $\epsilon_{a}$ and $\epsilon_{r}$, relative truncation tolerance $\epsilon_t$, and (optional) maximum rank $r_{max}$.

        \State $\delta_r \leftarrow \infty$.
        \State $\mathcal{S}^{n} \leftarrow \left(\mathbf{A}^{n}, p_e^{n}\right)$.
        \State $\epsilon \leftarrow \epsilon_a + \epsilon_r\left\lVert\mathbf{H}\left(\mathcal{S}^{n}\right)\right\rVert_2$
        \While{$\delta_r > \epsilon$}
            \State $\mathcal{S}^{n+\theta,(k)} \leftarrow (1-\theta)\mathcal{S}^{n+1,(k)} + \theta\mathcal{S}^{n}$. \Comment{Blend the state once; all terms below use it.}
            \State $\mathbf{B}^{n+\theta,(k)} \leftarrow (\nabla\times\mathbf{A}^{n+\theta,(k)})$, $\mathbf{J}^{n+\theta,(k)} \leftarrow \frac{1}{\mu_0}(\nabla\times\mathbf{B}^{n+\theta,(k)})$.
            \State $\mathbf{E}^{n+\theta,(k)} \leftarrow -\left(\mathbf{A}^{n+1,(k)} - \mathbf{A}^{n}\right)/\Delta t$, $\quad\mathbf{E}^{*,n+\theta,(k)} \leftarrow \mathbf{E}^{n+\theta,(k)}$. \Comment{Faraday gives the full field $\mathbf{E}$; $\eta=0$.}
            \State $f^{n,(k+1)} \leftarrow \text{VLASOV}(f^n,f^{n,(k)},\mathbf{E}^{*,n+\theta,(k)}, \mathbf{B}^{n+\theta,(k)}, \epsilon_t, r_{max})$. \label{alg-line:VLASOV-call}  \Comment{Described in Algorithm \ref{alg:CN_Vlasov}.}
            \State $f^{n+\theta,(k+1)} \leftarrow (1-\theta)f^{n,(k+1)} + \theta f^{n}$. \Comment{Blend $f$ as with $\mathcal{S}$.}
            \State $n^{n+\theta,(k+1)} \leftarrow \left\langle 1, f^{n+\theta,(k+1)}\right\rangle$, $(n\mathbf{u})^{n+\theta,(k+1)} \leftarrow \left\langle \mathbf{v}, f^{n+\theta,(k+1)}\right\rangle$, $\kappa^{n+\theta,(k+1)} \leftarrow \left\langle \frac{1}{2}v^2, f^{n+\theta,(k+1)}\right\rangle$.
            \State $\mathbf{u}^{n+\theta,(k+1)}_{e} = \mathbf{u}^{n+\theta,(k+1)} - \frac{1}{en^{n+\theta,(k+1)}}\mathbf{J}^{n+\theta,(k)}$.
            \State $\mathbf{E}^{n+\theta,(k+1)} \leftarrow -\left(\mathbf{u}^{n+\theta,(k+1)}\times\mathbf{B}^{n+\theta,(k)}\right) + \frac{1}{en^{n+\theta,(k+1)}}\left(\left(\mathbf{J}^{n+\theta,(k)}\times\mathbf{B}^{n+\theta,(k)}\right) - \nabla(p_e)^{n+\theta,(k)}\right)$. \Comment{Ohm's law; $\eta=0$.}
            \State $\mathbf{H}\left(\mathcal{S}^{n+1,(k)}\right)_\mathbf{A} \leftarrow \left(\mathbf{A}^{n+1,(k)} - \mathbf{A}^{n}\right)/\Delta t + \mathbf{E}^{n+\theta,(k+1)}$. \Comment{Induction residual uses the full field $\mathbf{E}$.}
            \State $\mathbf{H}\left(\mathcal{S}^{n+1,(k)}\right)_{p_e} \leftarrow \left(p_{e}^{n+1,(k)} - p_{e}^n\right)/\Delta t + \left(p_{e,adv}\right)^{n+\theta,(k)}$. \Comment{$p_{e,adv}$ computed via \eqref{eq:p_e_adv}.}
            \If{Lagrange Multiplier}
                \State Compute $R_i$ variables described in Appendix \ref{appendix:lagrange-multipliers}.
                \State $\mathbf{H}\left(\mathcal{S}^{n+1,(k)}\right)_{p_e} \leftarrow \mathbf{H}\left(\mathcal{S}^{n+1,(k)}\right)_{p_e} - \left(\gamma-1\right)\sum_{i=1}^5 R_i$.
            \EndIf
            \State $\delta \mathcal{S}^{n+1,(k)} \leftarrow \mathcal{S}^{n+1,(k)} - \Delta t\mathbf{H}\left(\mathcal{S}^{n+1,(k)}\right)$.
            \State Compute $\vec{\alpha}^{(k)}$ from Least Squares.
            \State $\mathcal{S}^{n+1,(k+1)} \leftarrow \sum_{i=k-m}^k\alpha_{i-(k-m)}\delta\mathcal{S}^{n+1,(i)}$.
            \State $\delta_r \leftarrow \left\lVert\mathbf{H}\left(\mathcal{S}^{n+1,(k+1)}\right)\right\rVert_2$.
            \State $k \leftarrow k+1$.
        \EndWhile
        \State $\mathcal{S}^{n} \leftarrow \mathcal{S}^{n+1,(k)}$, $f^{n} \leftarrow f^{n,(k)}$, $k \leftarrow 0$.
    \end{algorithmic}
\end{algorithm}

\begin{algorithm}
\caption{Outline of the Implicit Midpoint Vlasov Solve}\label{alg:CN_Vlasov}
    \begin{algorithmic}

        \State \textbf{Given}: Collection of $N_x\times N_y\times N_z$ $d$-dimensional tensors $(f^n_{ijk}(\mathbf{v}), f^{n,(k)}_{ijk}(\mathbf{v})) \rightarrow (f^n, f^{n,(k)})$, $N_x\times N_y \times N_z$ arrays $\left(\mathbf{E}^*_{ijk}, \mathbf{B}_{ijk}\right) \rightarrow \left(\mathbf{E}^*, \mathbf{B}\right)$, relative truncation tolerance $\epsilon_t$, and (optional) maximum rank $r_{max}$.

        \State $f^{tmp} \leftarrow \frac{1}{2}\left(f^{n,(k)} + f^{n}\right)$.
        \State Compute spatial gradient components via central difference or QUICK, $\left(\frac{\delta f}{\delta \eta}\right) \leftarrow \mathcal{D}_\eta\left[f^{tmp}\right], \eta \in \{x,y,z\}$.
        \State Compute velocity gradient components via central difference, $\left(\frac{\delta f}{\delta v_\eta}\right) \leftarrow \mathcal{D}_{v_\eta}\left[f^{tmp}\right], \eta \in \{x,y,z\}$.
        \If{Lagrange Multiplier}
            \State $\gamma_{v_\eta} \leftarrow -\left\langle f,1\right\rangle \big/ \left\langle v_\eta, \left(\frac{\delta f}{\delta v_\eta}\right)\right\rangle$.
            \State $\lambda_{v_\eta} \leftarrow \left(\left\langle v_\eta, -\gamma_{v_\eta}\left(\mathbf{u}\times\mathbf{B}\right)_\eta\left(\frac{\delta f}{\delta v_\eta}\right)\right\rangle + \left\langle v_\eta, \left(\mathbf{v}\times\mathbf{B}\right)_\eta\left(\frac{\delta f}{\delta v_\eta}\right)\right\rangle\right) \big/ \left\langle v_\eta, \left(\frac{\delta f}{\delta v_\eta}\right)\right\rangle$.
            \State $\Gamma \leftarrow \text{diag}(\gamma_{v_x}, \gamma_{v_y}, \gamma_{v_z})$, $\vec{\lambda} \leftarrow (\lambda_{v_x}, \lambda_{v_y}, \lambda_{v_z})^T$
            \State $\hat{\mathbf{E}}^* \leftarrow \Gamma\mathbf{E}^* - \vec{\lambda}$.
        \Else
            \State $\hat{\mathbf{E}}^* \leftarrow \mathbf{E}^*$.
        \EndIf
        \State $\text{ADV}_{x} \leftarrow \mathbf{v}\cdot\left(\frac{\delta f}{\delta \mathbf{x}}\right)$.
        \State $\text{ADV}_{v} \leftarrow \frac{q}{m}\left(\hat{\mathbf{E}}^* + \mathbf{v}\times\mathbf{B}\right)\cdot\left(\frac{\delta f}{\delta \mathbf{v}}\right)$.
        \State $f^{n,(k+1)} \leftarrow f^n - \Delta t\left(\text{ADV}_{x} + \text{ADV}_{v}\right)$
        \State Compute moments $\rho$, $\mathbf{J}$, $\kappa$ from $f^{n,(k+1)}$.
        \State $f^{n,(k+1)}_{t} \leftarrow \mathcal{T}_{\epsilon_t, r_{max}}\left[f^{n,(k+1)}\right]$. \Comment{Described in Section \ref{subsection:conserving-truncation}.}
        \State Compute moments $\rho_{t}$, $\mathbf{J}_{t}$, $\kappa_{t}$ from $f^{n,(k+1)}_{t}$; compute $K$ variables \eqref{eq:K_variables}-\eqref{eq:K_1_tilde_variable}; compute $f^{n,(k+1)}_{c}$ \eqref{eq:correction_tensor_rank_2}.
        \State $f^{n,(k+1)} \leftarrow f^{n,(k+1)}_{t} + f^{n,(k+1)}_{c}$.
    \end{algorithmic}
\end{algorithm}

A few remarks on notation are in order. We assume without loss of generality a single species distribution function at spatial point $(x_i,y_j,z_k)$ $f_{s,ijk}(\mathbf{v}) \rightarrow f(\mathbf{v})$, leaving out the $i,j,k$ subscripts to reduce tedium. We then let this $f(\mathbf{v})\rightarrow f$ denote a $d$-dimensional tensor at this spatial point. Algorithms \ref{alg:hybrid-method} and \ref{alg:CN_Vlasov} should be understood as performing their operations for each tensor object across the discretized physical domain. We also distinguish the rank ceiling $r_{max}$, an external limit we sometimes impose to bound the memory footprint, from the actual rank $r$, which we take to be the maximum over all core tensor ranks. Finally, while the $\mathbf{v}$ coordinates are identical for every tensor in our simulations, this need not hold in general; each tensor's velocity coordinates should be understood as having its own spatial location $i,j,k$ and $\Delta v$, which may itself be nonuniform. We will continue with our previous convention of letting $\eta \in \{x,y,z\}$. When considering moment weights, we recall that $1$, $v^2$, as well as the individual components of the velocity vector $\mathbf{v} = \left(v_x,v_y,v_z\right)$, are all $N_{v_x}\times N_{v_y} \times N_{v_z}$ arrays. We refer the reader to Section \ref{subsection:discrete-hybrid-model} for more details on this.

It should be emphasized that the conserving discrete scheme discussed in Sections \ref{subsection:discrete-hybrid-model} and \ref{section:conservative-update-scheme}, combined with the modified conserving truncation method described in Section \ref{subsection:conserving-truncation}, enables us to avoid solving separate fluid equations in order to conserve the moments. We simply update the distribution function, truncate, and compute the moments, assured that they will be globally conserved. Now, preservation of the moment conditions depends, in part, on $\epsilon$, which governs how tightly the implicit midpoint iteration must converge. Left too loosely converged, the nonlinear solve can leak spurious information into the moments, through no fault of the truncation method or kinetic update, but through the incompletely converged solve itself. This will be relevant to our numerical results, which we now turn to examine.

\section{Numerical Results} \label{section:numerical-results}

\begin{table}
    \centering
    \begin{tabular}{|c||c|c|c|}
        \hline
        Parameter & Landau Damped IAW & Whistler Wave & Proton Cyclotron Anisotropy Instability \\
        \hline
         $L_x$               			 & $16$         & $32$      & $10.5$    \\ 
         $N_x$               			 & $128$        & $64$      & $64$      \\ 
         $v_{span}$          			 & $9$          & $9$       & $9$       \\ 
         $N_v$               			 & $100$        & $100$     & $100$     \\ 
         $\Delta t$          			 & $10^{-2}$    & $10^{-2}$ & $10^{-2}$ \\
         $\epsilon_{vx}$     			 & $10^{-2}$    & $0$       & $0$       \\ 
         $\epsilon_{vy}$     			 & $0$          & $10^{-1}$ & $0$       \\ 
         $\epsilon_{vz}$     			 & $0$          & $10^{-1}$ & $10^{-7}$ \\ 
         $\epsilon_{Ay}$     			 & $0$          & $10^{-1}$ & $0$ \\ 
         $\epsilon_{Az}$     			 & $0$          & $10^{-1}$ & $10^{-7}$ \\ 
         $d_i$               			 & $1$          & $1$       & $1$ \\
         $B_0$               			 & $0$          & $1$       & $1$ \\
         $H_x$               			 & $1$          & $1$       & $1$ \\
         $k_x \equiv 2\pi H_x/L_x$ & $\pi/8$ 	    & $\pi/16$  & $4\pi/21$ \\
         $v_{th,x}$          			 & $\sqrt{1/3}$ & $1/5$     & $\sqrt{1/2}$ \\ 
         $v_{th,y}$          			 & $1.0$        & $1/5$     & $\sqrt{3/2}$ \\ 
         $v_{th,z}$          			 & $1.0$        & $1/5$     & $\sqrt{3/2}$ \\ 
         $\tau$              			 & $1/5$        & $1$       & $1$ \\ 
         $\gamma$            			 & $5/3$        & $5/3$     & $5/3$ \\ 
         \hline
    \end{tabular}
    \caption{The parameters for each problem setup.}
    \label{tab:all-problem-setups}
\end{table}

Our research code, written in Fortran, takes advantage of the Tensors Optimized for High-level Research (THOR) library \cite{alexandrov2024thor}, which provides a set of tensor objects and operators. Instead of storing all of phase space in a TT, we store a TT object representing the distribution function at each physical gridpoint. Said differently, we only compress velocity space. We consider three problems of varying complexity, each $1$D$3$V, as the multidimensional nature of the velocity component is sufficient to show the capabilities of our moment-conserving truncation method. The initial (unperturbed) moments for each problem may be found in Table \ref{tab:all-problem-setups}. Perturbations to the distribution function may be thought of as adding drift. We also eliminate all source terms from our problem, setting $\eta, \kappa_e$, and $\mu_e$ all to zero. We introduce a new physical parameter $d_i$ representing ion skin depth.

In all cases, each physical node is stored by way of a three-dimensional THOR TT object. The initial distribution function for each velocity dimension is a Maxwellian (which is notably rank 1)

\begin{equation}
    f_0(v_\eta) = \frac{n_0}{\sqrt{2\pi}\,v_{th,\eta}}\exp\left[-\left(v_\eta - v_{d,\eta}\right)^2/\left(2v_{th,\eta}^2\right)\right], \eta\in\{x,y,z\}.
\end{equation}

For our problems $n_0 = 1$ and $v_{d,\eta} = 0 \, \forall \, \eta$. The distribution function is discretized along discrete physical space $x_i$, $f_i(v_x,v_y,v_z) = f(x_i, v_x, v_y, v_z)$. $v_\eta$ is discretized into $N_{v,\eta}$ nodes (though all our problems set this to be uniform for all $\eta$), and in all problems has its boundaries at $\pm v_{span}v_{th,\eta}$ (see Section \ref{subsection:discrete-hybrid-model} for more details on discretization). At each point the tensor object is constructed $f_i(v_x,v_y,v_z) = f_i(v_x) \otimes f_i(v_y) \otimes f_i(v_z)$.

\subsection{Landau Damped Ion Acoustic Wave} \label{subsection:IAW}

The hybrid method of \cite{stanier2019fully} runs a $1$D$1$V Landau damping problem which we replicate here, expanding to $1$D$3$V.\footnote{For more details, Chapter 18, Sections 3 and 4 of Ref. \cite{Bittencourt} give a good discussion on deriving both the dispersion relation and Landau damping.} For our problem setup, we perturb as follows

\begin{equation}
    f(x,v_x,v_y,v_z,0) = \left(1 - \epsilon_{vx}\cos\left(k_xx\right) \frac{\partial}{\partial v_x}\right)f_0.
\end{equation}

The dispersion relation is $Z'(\xi) = 2\tau$, where $\tau \equiv \frac{T_i}{T_e}$ and $Z$ is the dispersion function or the scaled Faddeeva function $w$:
\begin{equation}
    Z(\xi) = \frac{1}{\pi^{1/2}}\int_{-\infty}^{+\infty}\frac{\exp(-q^2)}{q-\xi}dq \equiv i\sqrt{\pi}w(\xi).
\end{equation}

The plasma dispersion function follows the identity $Z'(\xi) = -2\left[1 + \xi Z(\xi)\right]$, which in turn implies
\begin{equation}
    1 + \xi Z(\xi) + \tau = 0.
\end{equation}

The Faddeeva function is well studied, and utilizing any numerical solver (in our case Matlab) yields the following solution:
\begin{equation}
    \xi = \frac{\omega + i\gamma}{k\sqrt{2}v_{th,x}}, \quad \omega = 0.6827 = 2.13k\sqrt{2}v_{th,x}, \quad \gamma = -0.05265 = -0.164 k\sqrt{2}v_{th,x}.
\end{equation}

Here $\gamma$ is the Landau damping factor (not to be confused with the ratio of specific heats $\gamma=5/3$). 

\begin{figure}
    \centering
    \includegraphics[width=1.0\linewidth]{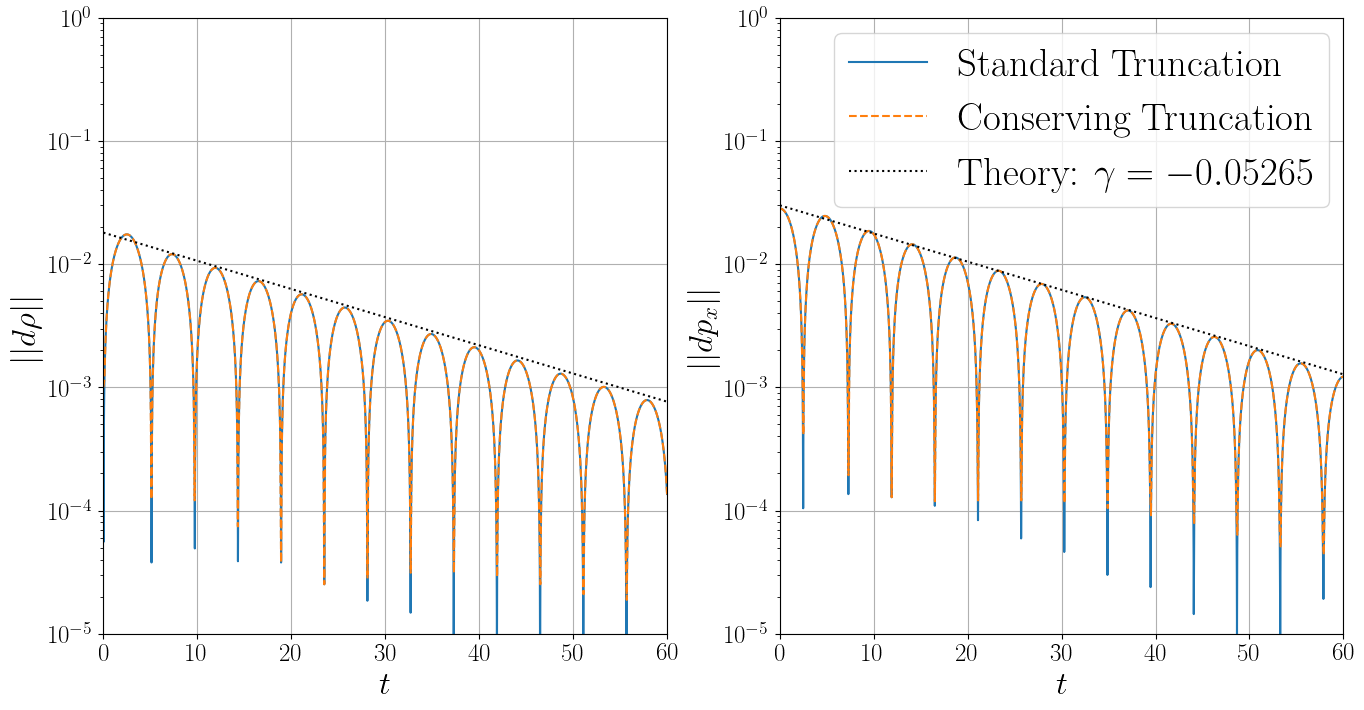}
    \caption{\textbf{Ion Acoustic Wave Decay Rate.} Perturbations of density (left) and momentum (right). Both follow the theoretical decay rate quite closely, regardless of truncation method. $\epsilon_{r}$ and $\epsilon_t$ are both $10^{-6}$.}
    \label{fig:iaw_damping_charge_momentum}
\end{figure}

The full details for the setup parameters for this problem may be found in Table \ref{tab:all-problem-setups}. For the results shown, the truncation tolerance, $\epsilon_t$ and the nonlinear relative tolerance, $\epsilon_{r}$, were both set to $10^{-6}$, though several variations of these tolerances were tried with negligible differences. We plot the $L^2$ norm of the density and momentum perturbation in Figure \ref{fig:iaw_damping_charge_momentum}. We see good agreement with the analytic decay rate with both the standard and the conserving truncation. Given the electric field is aligned in the $x$-direction (along the coordinate axis), the solution rank remains rank 1 in both cases; momentum and energy conservation errors remain at $\mathcal{O}\left(10^{-14}\right)$ (not shown).

\subsection{Whistler Wave} \label{subsection:whistler_numerical_results}

The equilibrium is a uniform plasma threaded by a constant background field $\mathbf{B}_0 = B_0\,\hat{\mathbf{x}}$ with $B_0 = 1$, aligned with the propagation direction $\mathbf{k} = k_x\hat{\mathbf{x}}$ (parallel propagation). The wave is introduced as a transverse perturbation of the vector potential about $\mathbf{A}_0 = \mathbf{0}$, so that $\nabla\times\mathbf{A}$ perturbs only the transverse components $B_y, B_z$ while $\mathbf{B}_0$ supplies the mean field.

The whistler wave may be set up as follows. We perturb the initial distribution function $f_0$ in the $v_y$ and $v_z$ directions
\begin{equation}
    f(x,v_x,v_y,v_z,0) = \left(1 + \frac{\epsilon_{vy}}{v_+}\cos\left(k_xx\right)\frac{\partial}{\partial v_y} - \frac{\epsilon_{vz}}{v_+}\sin\left(k_xx\right)\frac{\partial}{\partial v_z} \right)f_0, \quad v_+ \coloneqq \frac{1}{2}d_ik_x + \sqrt{1+\frac{1}{4}d_i^2k_x^2}.
\end{equation}

Additionally, we perturb the vector potential $\mathbf{A}$ as follows
\begin{equation}
    A_y(x) = A_{y,0} + \frac{\epsilon_{Ay}}{k_x}\cos\left(k_x x\right), \quad A_z(x) = A_{z,0} - \frac{\epsilon_{Az}}{k_x}\sin\left(k_x x\right), \quad A_{y,0} = A_{z,0} = 0
\end{equation}

Table \ref{tab:all-problem-setups} contains the parameter values. Under these conditions, we should see simple propagation of the $v_y$ and $v_z$ velocities along the $x$ direction \cite{stanier_2020}
\begin{align}
    v_y(x,t) &= -\frac{\epsilon_{vy}}{v_+}\cos(k_x\left(x - v_+ t\right)), \\
    v_z(x,t) &= \frac{\epsilon_{vz}}{v_+}\sin(k_x\left(x - v_+ t\right)).
\end{align}

\begin{figure}[ht]
    \centering
    \includegraphics[width=1.0\linewidth]{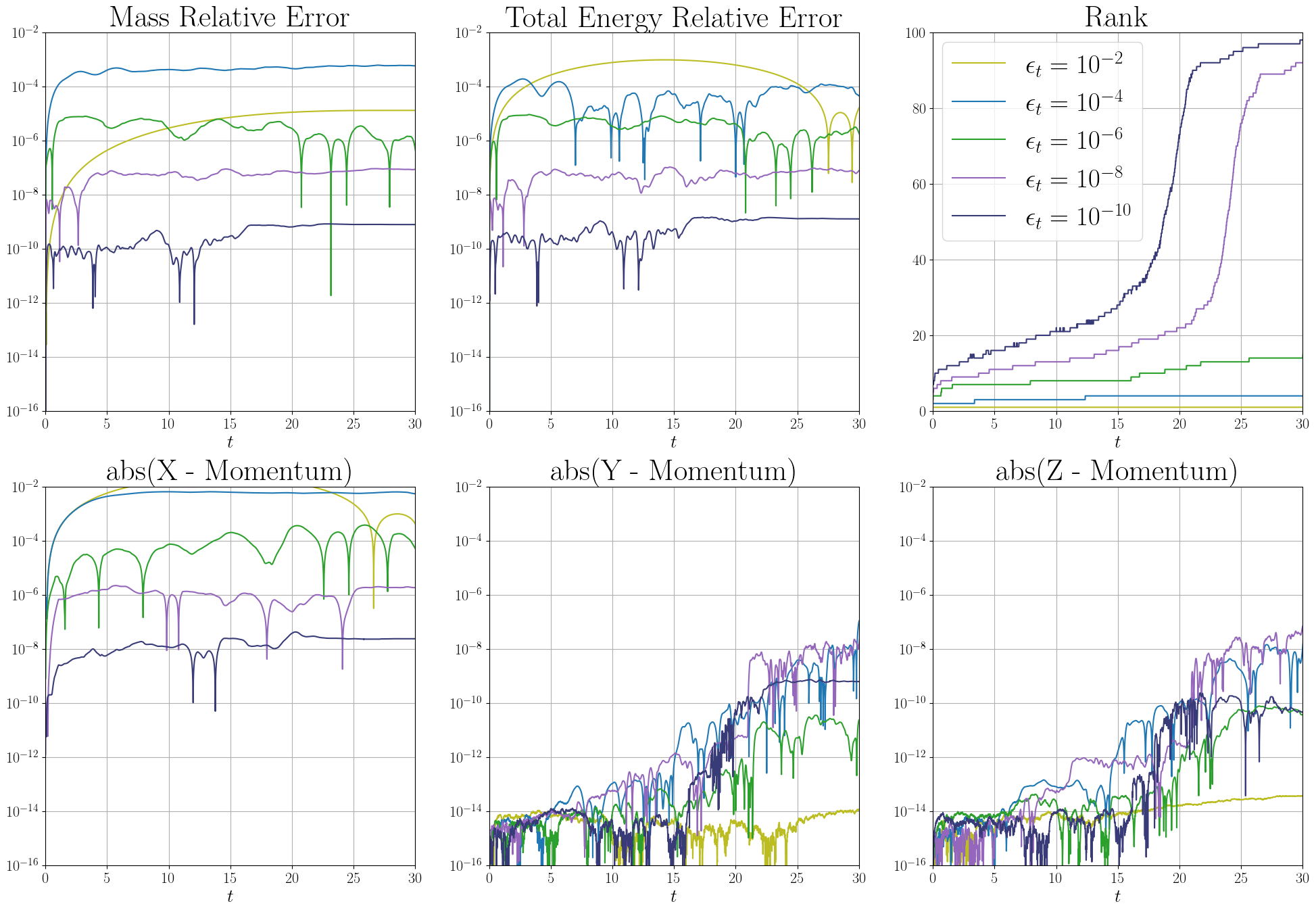}
    \caption{\textbf{Whistler Wave Moments.} Whistler wave moments refined over $\epsilon_t$ for $\epsilon_{r} = 10^{-10}$ using standard truncation.}
    \label{fig:whistler_all_moment_refinement_std}
\end{figure}

\begin{figure}[ht]
    \centering
    \includegraphics[width=1.0\linewidth]{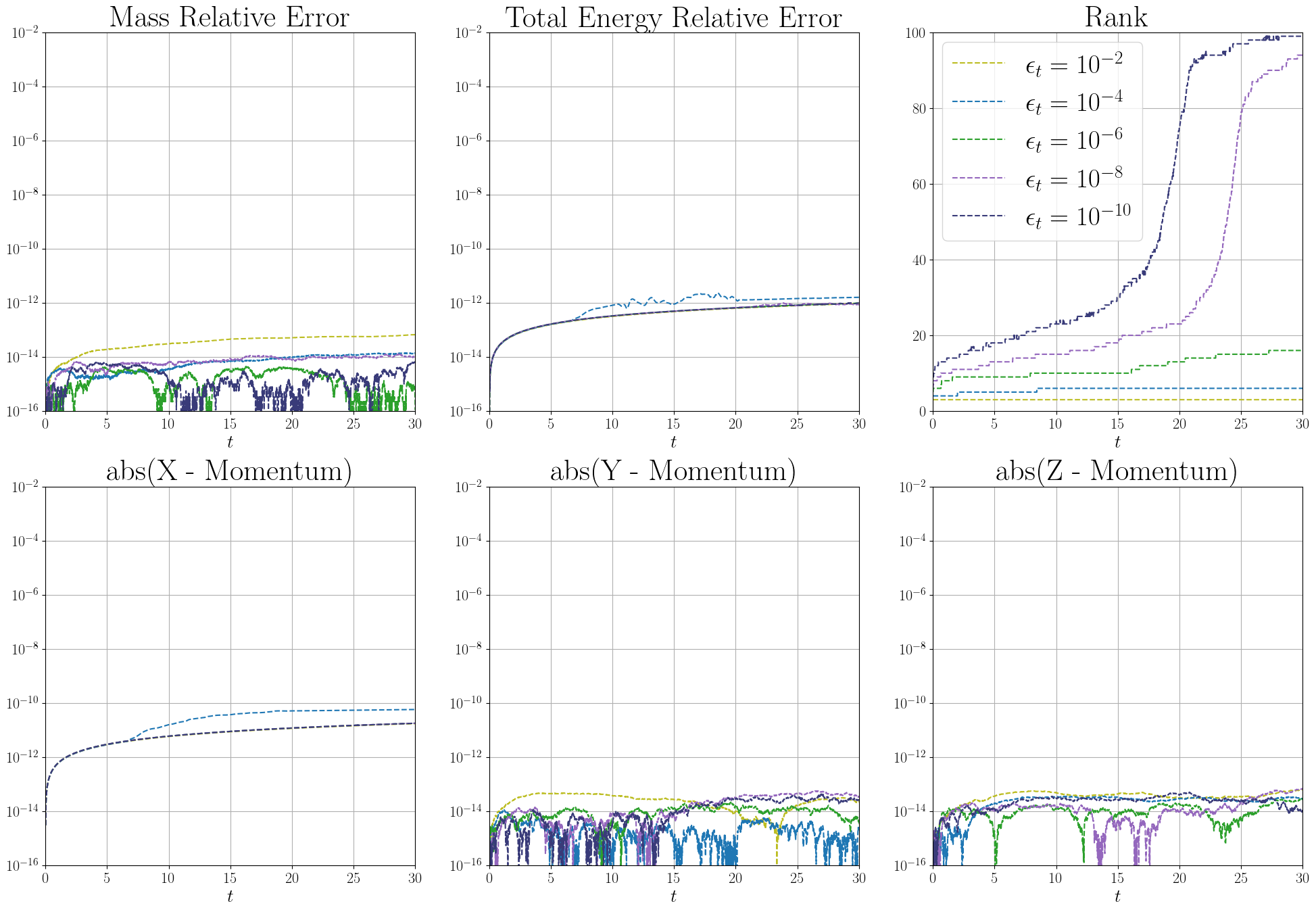}
    \caption{\textbf{Whistler Wave Moments.} Whistler wave moments refined over $\epsilon_t$ for $\epsilon_{r} = 10^{-10}$ using conservative truncation.}
    \label{fig:whistler_all_moment_refinement_cor}
\end{figure}

Interested in how the moments behave over a longer term and for a wide variety of truncation tolerances, we ran a sequence of runs with $\epsilon_t \in 10^{\{-2,-4,-6,-8,-10\}}$ and $\Delta t = 10^{-2}$ to a final time of $t=30$. We observe in Figures \ref{fig:whistler_all_moment_refinement_std} and \ref{fig:whistler_all_moment_refinement_cor} that the standard truncation method steadily improves in its moment relative errors as its truncation tolerance tightens, whereas we see negligible difference in any of the errors for the conserving method; all conservative truncation runs conserve mass, momentum, and energy to within $\epsilon_r = 10^{-10}$. Note how both methods drastically increase in rank the tighter the truncation tolerance. The Whistler wave's dispersion relation, $\omega = v_+ k_x$, becomes quadratic at high wavenumber, $\omega \propto k_x^2$, since $v_+ \to d_i k_x$. Thus, the dynamics increasingly emphasize high-wavenumber modes, precisely where the central difference stencil incurs its largest phase error and where its non-dissipative character leaves poorly resolved oscillatory content unattenuated. Driving $\epsilon_t$ well below the discretization error scale $\mathcal{O}(\max\{\Delta x^2,\Delta v^2\})$ causes the truncation to faithfully retain this fine-scale numerical content, which is poorly compressible in low rank and therefore drives steady rank growth without a corresponding gain in physical accuracy. If one is concerned about moment information, the loosest truncation tolerance with our conserving method conserves more information than the tightest standard method, yielding substantial savings in both memory footprint and runtime. 

\begin{figure}[ht]
    \centering
    \includegraphics[width=1.0\linewidth]{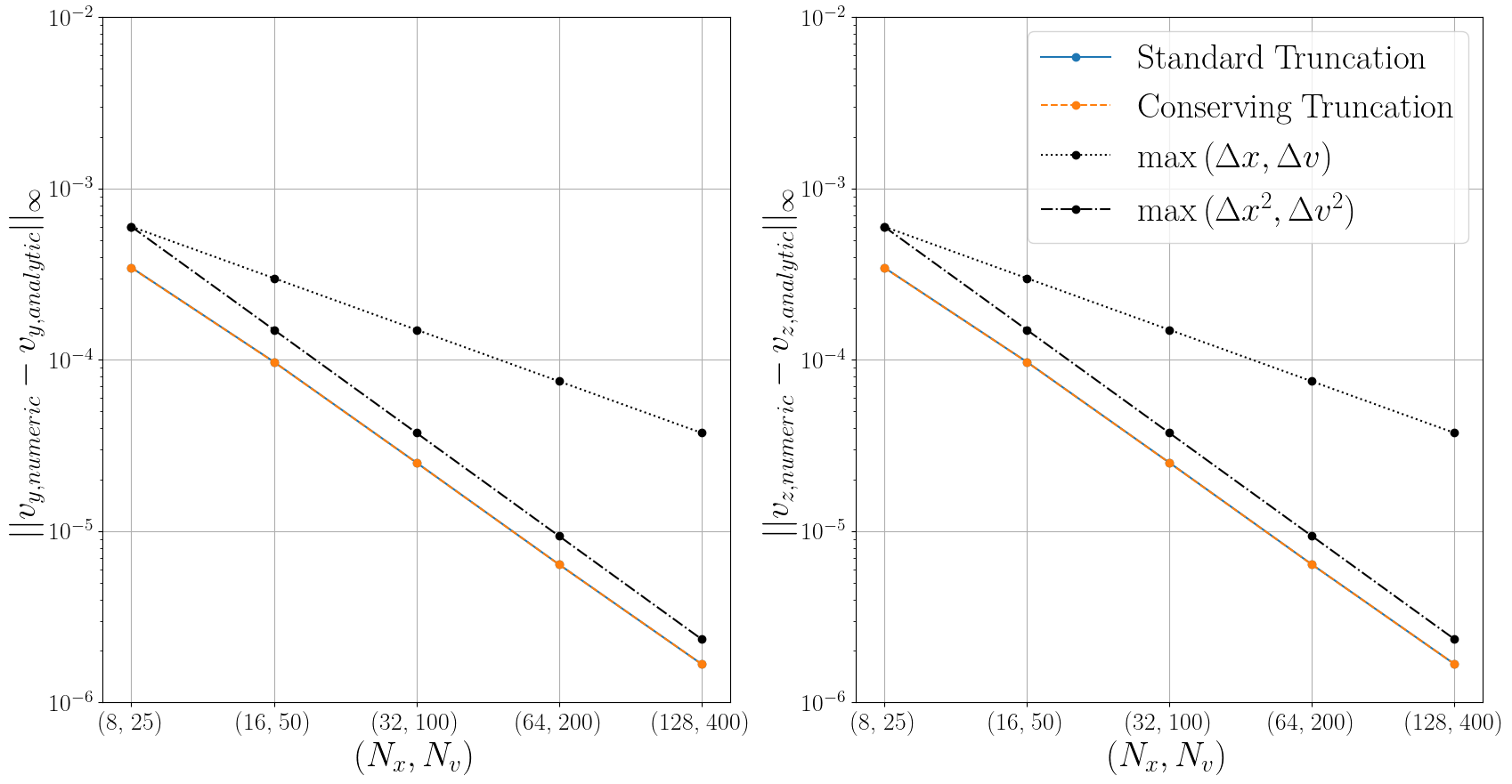}
    \caption{\textbf{Whistler Wave Refinement Study.} Setting $\Delta t = 10^{-5}$, and fixing a final time $t=10^{-1}$, we run the Whistler problem while refining by $\Delta x$ and $\Delta v$ and comparing against the analytic solution. Shown is with the tightest truncation tolerance we ran with, $\epsilon_t = 10^{-10}$, though the other tolerances behaved similarly.}
    \label{fig:whistler_refinement_study}
\end{figure}

Given the Whistler wave has analytic solutions, this also proved to be a good opportunity to perform a convergence study. We find our method, both with the standard and conserving truncation scheme, adheres quite closely to this analytic solution, as seen in Figure \ref{fig:whistler_refinement_study}. We tightened $\Delta t = 10^{-5}$ to prevent errors in time overwhelming errors in space or velocity, and then refined by $\Delta x$ and $\Delta v$, noting the quadratic refinement we would expect from a second-order method. Figure \ref{fig:whistler_refinement_study} displays the results for $\epsilon_t = 10^{-10}$, though it should be noted the results for other truncation tolerances were similar, with the standard truncation scheme and our conserving truncation scheme behaving the same.

\subsection{Proton Cyclotron Anisotropy Instability} \label{subsection:icai}

Given a static background magnetic field $\mathbf{B}_0$, a plasma with ion temperature ratio $T_{\perp}/T_{\parallel} = 3 > 1$, where $\perp$, $\parallel$ denote the relation to the magnetic field line, will produce an electromagnetic instability. Given a wave vector $\mathbf{k}$, the instability mode has maximum growth rate at $\mathbf{k}\times\mathbf{B}_0 = \mathbf{0}$. If we set both $\mathbf{k}$ and $\mathbf{B}_0$ along the $x$ axis, with the uniform background field $\mathbf{B}_0 = B_0\hat{\mathbf{x}}$, $B_0 = 1$, perturbing as follows gives us a $1$D$3$V electromagnetic instability to study:

\begin{align}
    f(x,v_x,v_y,v_z,0) &= \left(1 + \frac{\epsilon_{vz}}{v_+}\sin\left(k_xx\right) \frac{\partial}{\partial v_z}\right)f_0, \quad v_+ \coloneqq -\frac{1}{2}k_x + \sqrt{1 + \frac{1}{4}k_x^2}, \\
    A_z &= A_{z,0} - \frac{\epsilon_{Az}}{k_x}\sin\left(k_xx\right), \quad A_{z,0} = 0
\end{align}
\cite{Told_etal_2016} gives the dispersion relation for this problem in the context of the hybrid model, and using their code, HYDROS, we acquire the growth rate for the Proton Cyclotron Anisotropy Instability under the hybrid model (see also \cite{stanier2019fully}):
\begin{equation*}
    \gamma = 0.162.
\end{equation*}

\begin{figure}[ht]
    \centering
    \includegraphics[width=0.75\linewidth]{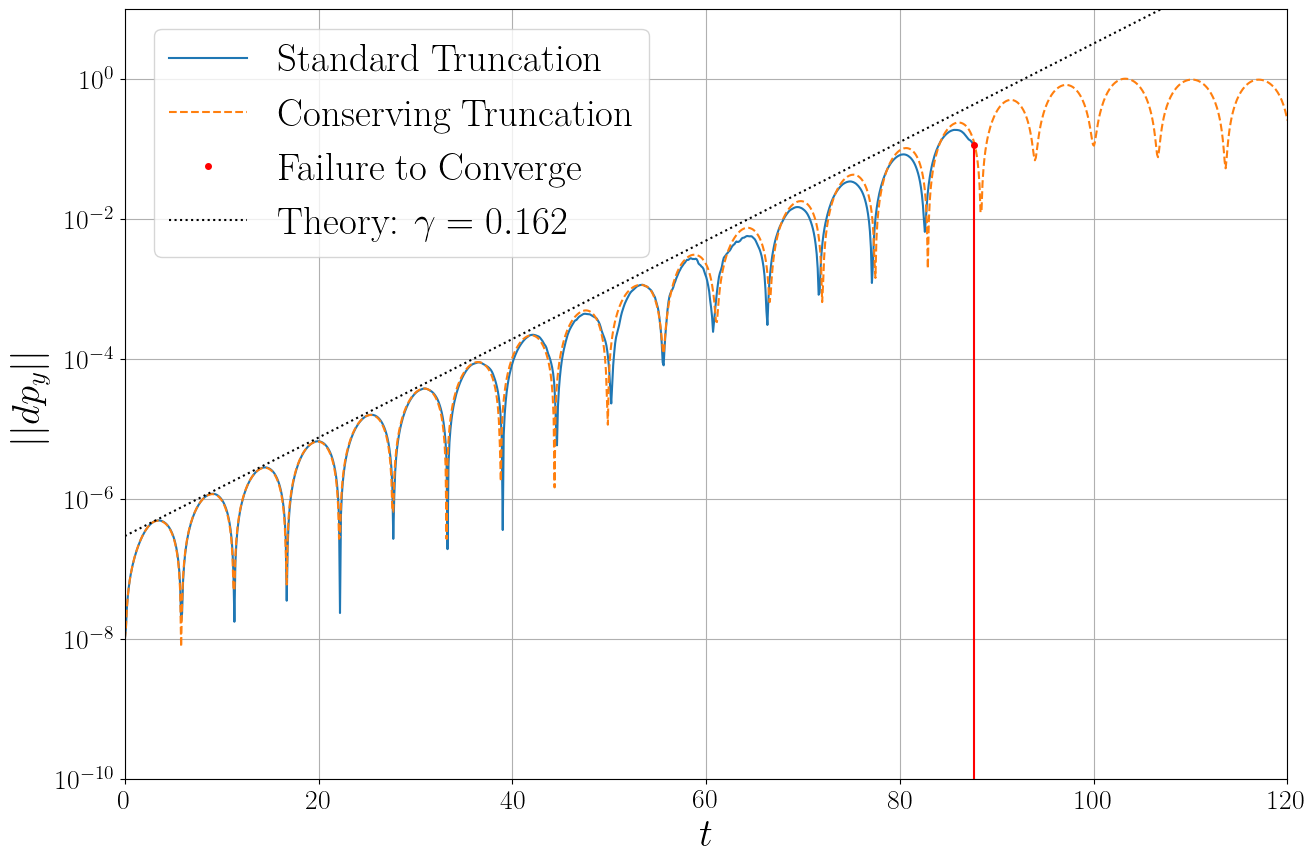}
    \caption{\textbf{Proton Cyclotron Anisotropy Instability Growth Rate.} The growth rate of the perturbation of $y$ momentum under both the standard truncation scheme and the conserving truncation scheme for the Proton Cyclotron Anisotropy Instability with $\epsilon_t = \epsilon_{r} = 10^{-10}$. Both match the theoretical growth rate obtained by HYDROS \cite{Told_etal_2016} until nonlinear effects take over. The run with standard truncation fails to converge at approximately $t=87$.}
    \label{fig:icai_growth_rate}
\end{figure}

\begin{figure}[ht]
    \centering
    \includegraphics[width=0.75\linewidth]{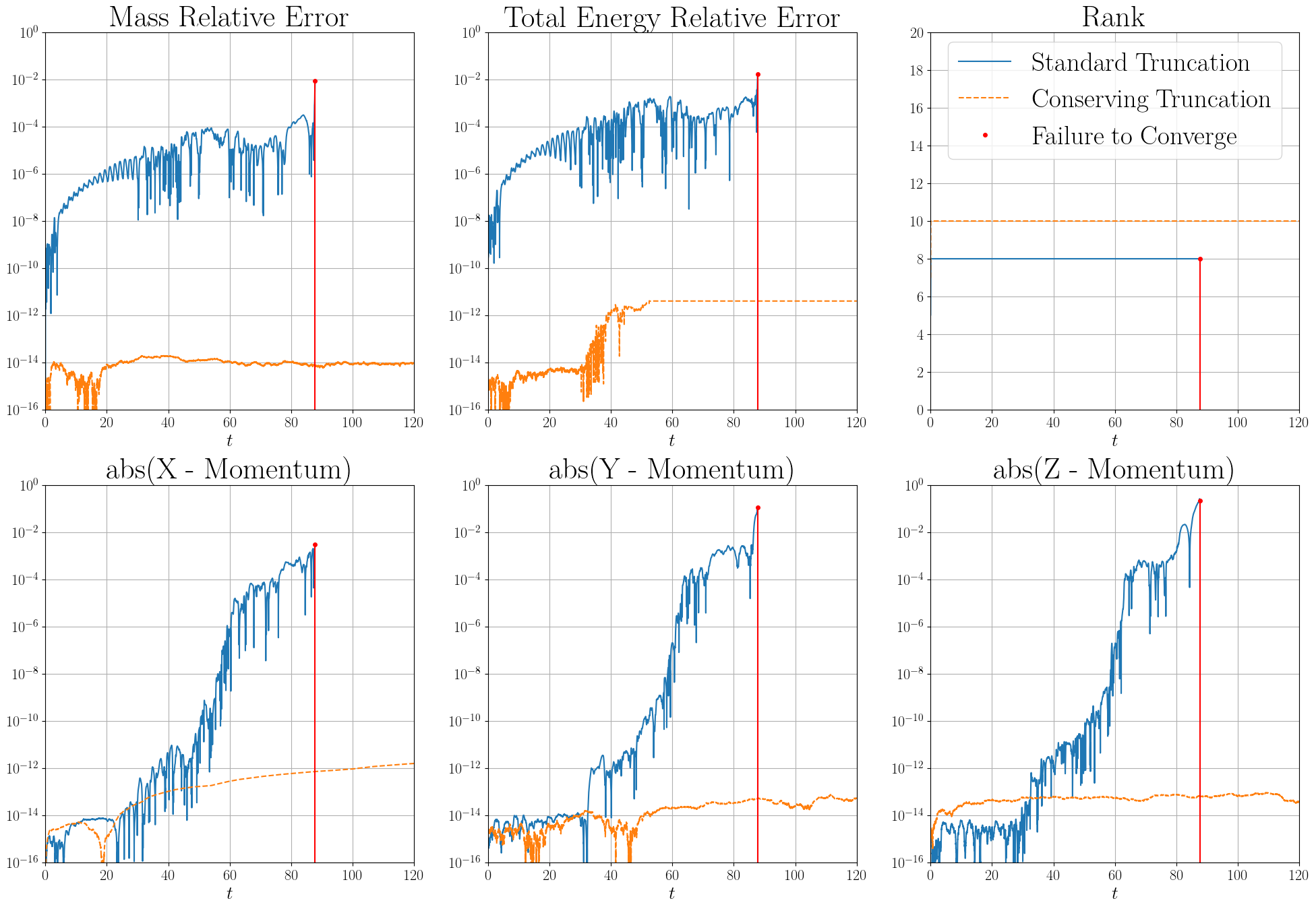}
    \caption{\textbf{Proton Cyclotron Anisotropy Instability Moments.} The relative error in the moments under both the standard truncation scheme and the conserving truncation scheme for the Proton Cyclotron Anisotropy Instability with $\epsilon_t = \epsilon_{r} = 10^{-10}$.}
    \label{fig:icai_moments}
\end{figure}

And indeed, we see in Figure \ref{fig:icai_growth_rate} our method simulates this problem accurately, following the growth rate until nonlinear effects take over the linear dispersion relation. Figure \ref{fig:icai_moments} shows preservation of the moments up to $\epsilon_{r} = 10^{-10}$. The truncation tolerance $\epsilon_t$ was set to $10^{-10}$, as a looser tolerance threatened the small perturbation associated with this problem. To reduce noise buildup, a tight cap was placed on the rank, with $r_{max} = 10$ for the conserving truncation scheme, and $r_{max} = 8$ for the standard,\footnote{Strictly speaking both had $r_{max}=8$, but the conserving truncation scheme adds $2$ rank in the correction step.} which means we may be cutting singular values substantially larger than $\epsilon_t$. This maximum rank results in the moment errors not being proportional to the truncation tolerance in the standard truncation case, unlike the Whistler moments in Figure \ref{fig:whistler_all_moment_refinement_std}.  

Readers will observe two runs in Figures \ref{fig:icai_growth_rate} and \ref{fig:icai_moments}. It was observed that, despite following the growth rate well, the standard truncation scheme failed to converge to a solution around $t=87$, unlike the conserving scheme. In order to run the code to completion using standard truncation, we needed to refine the mesh to $N_v=200$ and $\Delta t = 0.005$, significantly increasing the computational cost, which the conserving scheme was able to avoid.

See Table \ref{tab:all-problem-setups} for a full list of the problem parameters.

\subsection{Performance}

Some consideration of our method's performance is in order. We compare our implicit midpoint (IM) update scheme to two explicit Runge--Kutta update schemes (explicit midpoint and Kutta's classical third-order method, RK2 and RK3 respectively) with the following Butcher tableaux:

\begin{equation}
    \text{2-Stage Runge--Kutta:}\quad
    \begin{array}{c|cc}
        0   & 0   & 0 \\
        1/2 & 1/2 & 0 \\
        \hline
            & 0   & 1
    \end{array}
    \qquad\qquad
    \text{3-Stage Runge--Kutta:}\quad
    \begin{array}{c|ccc}
        0   & 0   & 0 & 0 \\
        1/2 & 1/2 & 0 & 0 \\
        1   & -1  & 2 & 0 \\
        \hline
            & 1/6 & 2/3 & 1/6
    \end{array}
\end{equation}

Both of these are explicit updates and therefore subject to a CFL restriction. 

\begin{figure}[ht]
    \centering
    \includegraphics[width=0.75\linewidth]{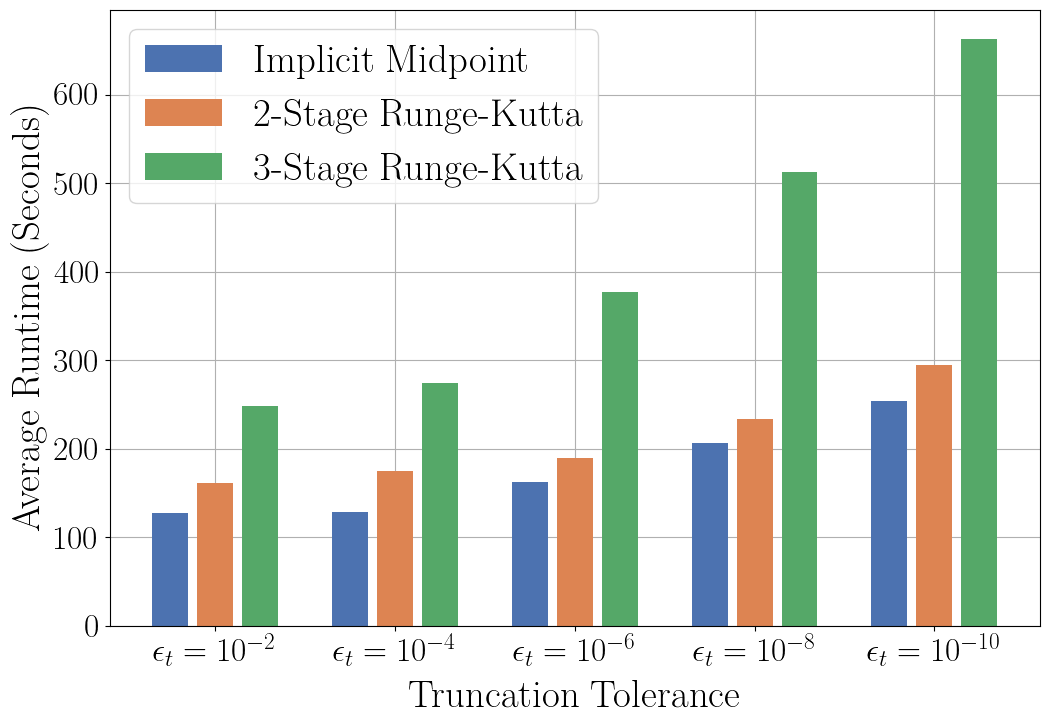}
    \caption{\textbf{Whistler Wave Walltime.} The average runtime of ten Whistler Wave simulations running to $t=1$ under implicit midpoint as well as 2- and 3-stage Runge-Kutta. Standard truncation calls were used in all cases, but similar performance behavior is seen when employing the conserving truncation scheme as well.}
    \label{fig:whistler_walltime_mp_intermediate_truncation}
\end{figure}

We use the Whistler Wave problem discussed in Section \ref{subsection:whistler_numerical_results}, running ten simulations to $t=1$. We observe both Runge--Kutta schemes slightly outperform our implicit midpoint rule if one looks strictly at the number of iterations in Algorithm \ref{alg:hybrid-method} it takes to converge, with both taking five iterations to the implicit midpoint's six. However, this is not the full picture, as observing the runtimes shows that, despite requiring more iterations to converge, the overall walltime of the implicit midpoint is considerably less (Figure \ref{fig:whistler_walltime_mp_intermediate_truncation}). This is due to the nature of the Runge--Kutta methods, the two- and three-stage methods effectively having respectively double and triple the number of operations to perform per non-linear iteration.

It is also worth briefly discussing optimization for this method. A simple truncation step at the end will certainly keep the rank growth in check throughout the simulation; however, for the Vlasov update itself this may not be sufficient. Recall from Section \ref{section:tensor-trains}, Table \ref{tab:tensor-operations-computational-complexity}, that tensor operations increase tensor rank. This results in a tensor with a rank that quickly balloons within Algorithm \ref{alg:CN_Vlasov}, called by Line \ref{alg-line:VLASOV-call} of Algorithm \ref{alg:hybrid-method}. While truncating at the end of the method will check this, it can result in significant increases in runtime, especially for a multistage method such as the two- or three-stage Runge--Kutta methods. The conservative truncation scheme is somewhat vulnerable to this as well, as it uses dot products to compute moments, whose runtime is $\mathcal{O}\left(d\,n\,r_\mathcal{A}r_\mathcal{B}\left(r_{\mathcal{A}} + r_{\mathcal{B}}\right)\right)$. See Tables \ref{tab:tt_op_counts} and \ref{tab:tt_op_extras} for more details on operation count for each method.

 \begin{table}[htbp]
    \centering
    \begin{tabular}{lccc}
      \toprule
      Operation & IM & RK2 & RK3 \\
      \midrule
      Tensor $\pm$ tensor$^{*}$        & 8+1   & 16 & 27 \\
      Scalar $\times$ tensor$^{*}$     & 8+1   & 16 & 30 \\
      Tensor $\star$ tensor            & 4     & 8  & 12 \\
      Scalar $+$ tensor                & 3     & 6  & 9  \\
      \bottomrule
    \end{tabular}

    \vspace{0.5em}

    {\footnotesize $^{*}$The initial averaging $f^{avg} = \frac{1}{2}\left(f^n + f^{n,(k)}\right)$ gives the IM method an additional tensor/tensor addition and scalar/tensor multiplication.\par}

    \caption{Tensor operations per spatial cell, per species, per time step. We assume $1$d$3$v, central differences in both $x$ and $v$, no Lagrange multiplier, standard truncation, and no $\epsilon_I$ truncation. The advection terms are evaluated once per stage, so the $2$- and $3$-stage Runge--Kutta methods perform roughly two and three times as many operations as the implicit midpoint method; RK3 incurs additional operations from combining its stages. Adding an $\epsilon_I$ truncation complicates the process, incurring additional operations when adding these advection terms.}
    \label{tab:tt_op_counts}
  \end{table}
  
    \begin{table}[htbp]
    \centering
    \begin{tabular}{lccccc}
      \toprule
      & Conserving & Lagrange & QUICK & \multicolumn{2}{c}{Extra spatial dim.} \\
      \cmidrule(lr){5-6}
      Operation & truncation & multiplier & diff. scheme & Central & QUICK \\
      \midrule
      Tensor $\pm$ tensor               & 5  & 3 & 10 & 2 & 12 \\
      Scalar $\times$ tensor            & 5  & 6 & 12 & 1 & 13 \\
      Tensor $\star$ tensor             & 0  & 3 & 3  & 1 & 4  \\
      Scalar $+$ tensor                 & 0  & 3 & 0  & 0 & 0  \\
      Dot product                       & 14 & 7 & 0  & 0 & 0  \\
      \bottomrule
    \end{tabular}

    \vspace{0.5em}
    \caption{Additional tensor operations under other settings. Conserving truncation operations are incurred per call to the truncation routine; all others are incurred per stage, making the $2$- and $3$-stage Runge--Kutta methods execute two and three times as many as the implicit midpoint method, respectively.}
    \label{tab:tt_op_extras}
  \end{table}

There are two ways we addressed this. The first is a relatively simple standard truncation applied with $\epsilon_{mp} = 10^{-14}$ after every tensor operation, cleaning up any noise that comes from the operation. We found this has no discernible changes to accuracy or moment preservation, as it is well beneath the nonlinear tolerance $\epsilon_r$. The second is adding an intermediate truncation step after adding every advection term. This intermediate truncation step is either conserving or standard, following the nature of the truncation method at the end of the Vlasov solver. We set $\epsilon_I = \epsilon_r$; since $\epsilon_r \leq \epsilon_t$, it has negligible impact on the moments in either the conserving or standard case. For the Runge--Kutta methods this may be further improved by adding a truncation step at the end of each stage rather than at the completion of the update, with this end-stage truncation method being conserving or standard, as the case may be.

Both of these truncations were found to improve runtime in most cases. The only exceptions to this were in cases where the rank of the tensors remained small, in which case the truncation operations cost more time than they saved. This only happens in the case of very loose $\epsilon_t$, and may be improved upon by adding rank checks before deciding to truncate.

The $\epsilon_{mp}$ truncation particularly benefits the implicit midpoint method. Algorithm \ref{alg:CN_Vlasov} starts by assigning $f^{tmp} \leftarrow \frac{1}{2}\left(f^n + f^{n,(k)}\right)$. Assuming each have rank $r$, $f^{tmp}$ will now have $2r$, given tensor addition concatenates cores (see Section \ref{section:tensor-trains}). However, the cores of each span nearly identical subspaces; it is in their coefficients that they differ. In fact, on the first nonlinear iteration $f^{n,(0)} = f^n$, so $f^{tmp}$ having rank of $2r$ implies that $r$ of its basis vectors are redundant. On subsequent iterations $f^{n,(k)} - f^n$ is an $\mathcal{O}(\Delta t)$ correction which, in our experiment, lies almost entirely within the existing subspace. It follows that a standard truncation of $f^{tmp}$ with $\epsilon_{mp}$ will return its rank to $r$ or close to it, removing redundancy rather than information. Without this step, a tensor of rank $2r$ rather than $r$ is passed through a battery of tensor operations all of which have costs that grow polynomially in rank. The Runge--Kutta methods do no such averaging, but they still benefit from removing potential redundant rank that comes with any tensor operation.

\begin{figure}[ht]
    \centering
    \includegraphics[width=0.75\linewidth]{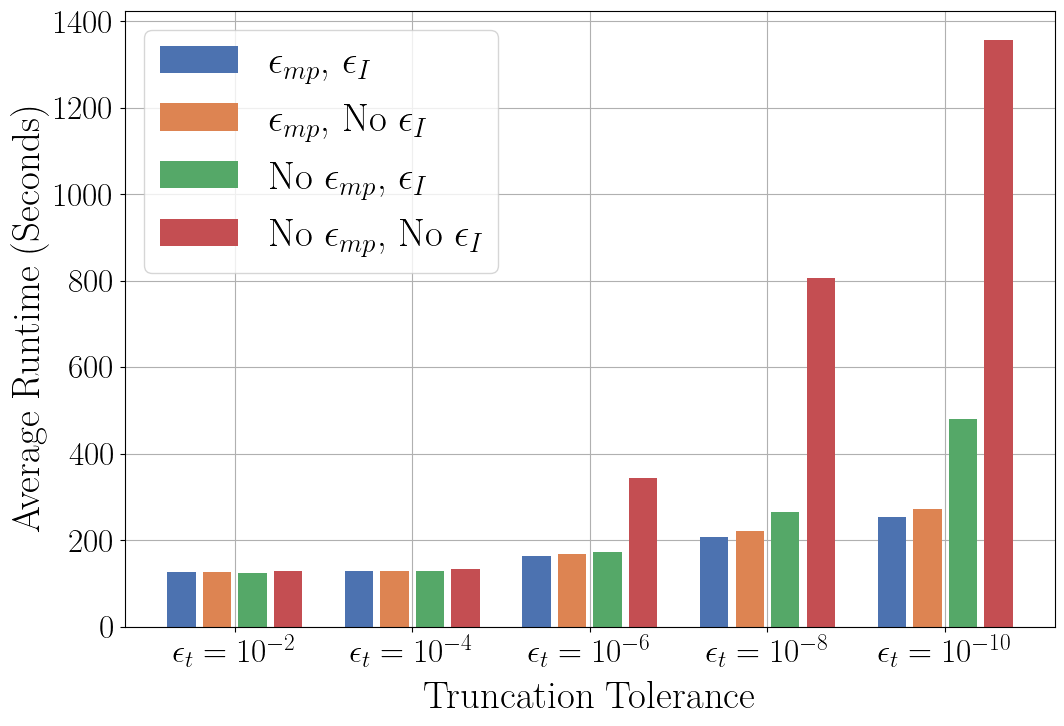}
    \caption{\textbf{Whistler Wave Walltime.} The average runtime of ten Whistler Wave simulations to $t=1$ under the implicit midpoint scheme (Algorithm \ref{alg:CN_Vlasov}). Standard truncation calls were used in all cases, but similar performance behavior is seen when employing the conserving truncation scheme as well.}
    \label{fig:whistler_walltime_auxiliary_truncation}
\end{figure}

The effect of both of these auxiliary truncation steps may be found in Figure \ref{fig:whistler_walltime_auxiliary_truncation}. It is clear that the $\epsilon_{mp}$ has the greatest effect, though both truncations only begin to have appreciable impact with tighter truncation tolerances.

\begin{figure}[ht]
    \centering
    \includegraphics[width=0.75\linewidth]{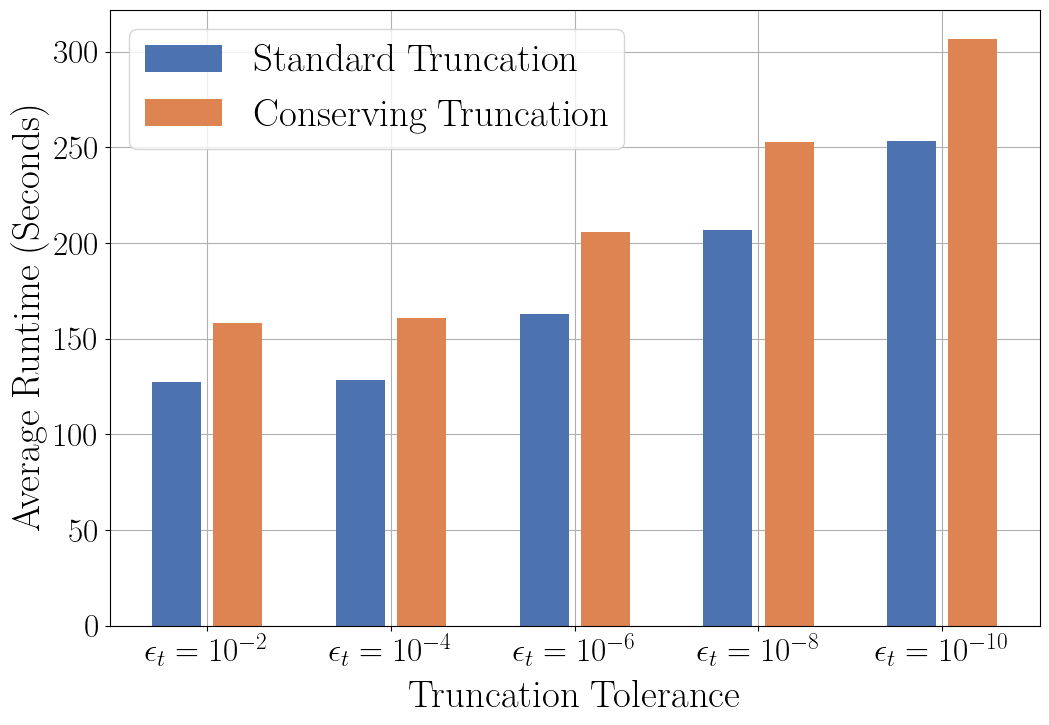}
    \caption{\textbf{Whistler Wave Walltime.} The average runtime of ten Whistler Wave simulations to $t=1$ under the implicit midpoint scheme (Algorithm \ref{alg:CN_Vlasov}) with both $\epsilon_I$ and $\epsilon_{mp}$ auxiliary truncations. We compare the performance of the standard truncation to the conserving.}
    \label{fig:whistler_walltime_truncation_types}
\end{figure}

Finally, we consider the performance between the standard and conserving truncation schemes. Figure \ref{fig:whistler_walltime_truncation_types} shows the same setup as we have been discussing, comparing the implicit midpoint update scheme using the standard and conserving truncation schemes. Not surprisingly, there is some runtime penalty due to the additional operations that come with the conserving truncation scheme (see Table \ref{tab:tt_op_extras}). The increase in runtime is consistently 10\%-20\% across the range of truncation tolerances, as such is not completely negligible; however, runtimes remain the same order of magnitude. It was discussed in Section \ref{subsection:icai} how this new truncation technique enables a coarser mesh, mitigating this cost. Section \ref{subsection:whistler_numerical_results} also discusses how a looser $\epsilon_t$ enables a lower rank, speeding the overall simulation while keeping the key moment information. In short, while the conserving truncation does, strictly speaking, take longer than the standard, it unlocks faster simulations on the whole.

\section{Conclusion} \label{section:conclusion}

We have presented a novel fully implicit, moment-conserving, electromagnetic, hybrid kinetic-ion/fluid-electron method. This method handles the Eulerian kinetic-ion distribution function via tensors, utilizing the Tensor Train decomposition to sidestep the Curse of Dimensionality. This method uses the implicit midpoint method for updating the field and fluid quantities while enslaving the kinetic distribution function to follow. It has been proven that the discrete method conserves mass, momentum, and energy up to the nonlinear tolerance. Though a standard truncation method of the distribution function will disrupt this conservation, this problem is solved by our improved conservative truncation method. When put together, this gives us a method that conserves the moments to nonlinear tolerance without needing to advance an auxiliary set of fluid equations for the moments. The physical accuracy and conservation of moments were demonstrated in a sequence of electrostatic and electromagnetic problems. The physics was shown to be preserved in the Ion Acoustic Wave and Whistler Wave, as well as the Proton Cyclotron Anisotropy Instability, while moment conservation held regardless of truncation tolerance, unlike the standard truncation, for the latter two.

A number of avenues follow naturally from the present work. First, the truncation tolerance $\epsilon_t$ and maximum rank $r_{max}$ are presently global parameters; an adaptive, local rank-selection criterion, driven by the local physics rather than a single global tolerance, would permit high rank only where the distribution genuinely demands it. Relatedly, the adaptive weighting scheme of Ref. \cite{zheng_sands_qiu_2026} offers a promising route to a more flexible conserving projection. Second, the incorporation of a collision operator, which drives the distribution function toward a rank-one Maxwellian, is expected to act as a natural rank-reducing mechanism, and its interplay with the low-rank structure warrants study. Third, while we compress only velocity space here, extending the Tensor Train representation to the configuration-space coordinates as well is the natural path toward the full-dimensional regimes that motivate this work. Finally, the field-fluid solve is presently unpreconditioned; the development of a physics-based preconditioner for the hybrid system would substantially improve the efficiency of the nonlinear solve.

\section*{Acknowledgments}
This work was supported by the U.S. Department of Energy through the Los Alamos National Laboratory. Los Alamos National Laboratory is operated by Triad National Security, LLC, for the National Nuclear Security Administration of U.S. Department of Energy (Contract No. 89233218CNA000001). Research presented in this article was supported by the Laboratory Directed Research and Development program of Los Alamos National Laboratory under project number 20250303ER.

\section*{Data Availability}
The data supporting the findings of this study are available from the corresponding author upon reasonable request.

\section*{Declaration of Competing Interest}
The authors declare that they have no known competing interests that are relevant to the work reported in this paper.

\appendix
\section{Lagrange Multipliers}\label{appendix:lagrange-multipliers}

Section \ref{section:conservative-update-scheme} proved our method conserves mass, momentum, and energy. However, if we assume a finite velocity domain, as is the case in any discrete code, momentum and energy conservation may fall apart, mostly due to errors that come in through the boundaries, which are supposed to extend to infinity. These proofs are predicated on the identity \eqref{eq:IBP-velocity}, rewritten here for convenience
\begin{equation} \label{eq:IBP-continuous-velocity-appendix}
    \int\chi\left(\mathbf{F}\cdot\nabla_v f\right)d\mathbf{v} = \cancel{\int_{\Omega_v}\nabla_v\cdot\left(\mathbf{F}\chi f\right)d\mathbf{v}} - \int_{\Omega_v}f\mathbf{F}\cdot\nabla_v\chi d\mathbf{v} - \cancel{\int_{\Omega_v}f\chi\nabla_v\cdot\mathbf{F}d\mathbf{v}} = -\int_{\Omega_v}f\mathbf{F}\cdot\nabla_v\chi d\mathbf{v}.
\end{equation}

Its discrete analog, reading $\nabla_v$ as a vector of discrete central difference operators $(D_\alpha f)_j = (f_{j+1}-f_{j-1})/(2\Delta v)$, is written
\begin{equation} \label{eq:IBP-discrete-velocity-appendix}
    \left\langle\chi,\mathbf{F}\cdot\nabla_v f\right\rangle = \cancel{\left\langle 1, \nabla_v\cdot\left(\mathbf{F}\chi f\right)\right\rangle} - \left\langle f, \mathbf{F}\cdot\nabla_v\chi\right\rangle - \cancel{\left\langle f\chi, \nabla_v\cdot\mathbf{F}\right\rangle}.
\end{equation}

This identity is predicated on velocity space extending to infinity, with $\lim_{v\rightarrow \pm \infty}f = 0$; if it does not, the first integral, $\int_{\Omega_v}\nabla_v\cdot\left(\mathbf{F}\chi f\right)d\mathbf{v}$, does not go to zero, but instead has a remainder that is equal to the sum of the boundary fluxes. The last integral goes to zero regardless, as $\nabla_v\cdot\mathbf{F} = 0$ for the Lorentz force.

The following will consider the moments applied to some single-species distribution function $f$. Generalizing to multiple species is intuitive enough, but the symbolic manipulation would be more involved. As throughout the rest of the paper, we neglect the $\overset\leftrightarrow{\Pi}_e$, $\mathbf{q}_e$, and $H_e$ source terms, effectively setting $\kappa_e$ and $\mu_e$ to zero, and we set $\eta$ to zero as well, i.e. $\mathbf{E} = \mathbf{E}^*$. Our test problems do not have these source terms, and the logic remains the same for correcting for any errors that these may bring in the discrete setting. Additionally, to reduce tedium, we forgo using spatial indices except where we explicitly are summing over the spatial domain. In what follows, all quantities are defined at cells unless otherwise stated. The $\nabla_x$ and $\nabla_v$ operators should be viewed as discrete nabla operators. The velocity derivatives are always central, the spatial derivatives are always central except in the case of the advective fluxes, which may be central or QUICK.

Finally, it is worth recalling that none of these errors are due to the nature of the truncation scheme nor to the implicit update, but rather from the nature of dealing with a finite mesh that has been discretized.

\subsection{Mass} \label{subsection:lagrange-multipliers-mass}

The acceleration operator of the Vlasov equation \eqref{eq:vlasov-s} goes to zero under the zeroth moment if we choose proper velocity boundary conditions. In the momentum and energy sections that follow, appeals to integration by parts will fall victim to the errors described above. However, in the special case where $\chi$ is not a function of $\mathbf{v}$, as is the case for the zeroth moment, one may appeal to integration by parts, that $\left\langle\mathbf{F},\nabla_v f\right\rangle = \left\langle 1, \nabla_v\cdot\left(f\mathbf{F}\right)\right\rangle-\left\langle f,\nabla_v\cdot\mathbf{F}\right\rangle$. The second term goes to zero due to $\nabla_v\cdot\mathbf{F}=0$. By appealing to the divergence theorem, we see the first term is the sum of the flux through the boundary faces. Of course, as $\mathbf{v}\rightarrow\infty$, this goes to zero, but even in the discrete setting on a finite domain, so long as we enforce zero flux through the boundaries, which may be done by setting antisymmetric boundary conditions when computing the $v$-advection term, mass will be conserved. This may be more clearly seen by the summation by parts telescoping sum:
\begin{align}
    \begin{split}
        \sum_{j=1}^{N-1} (f_{j+1} - f_{j-1}) &= f_2 - f_0 + f_3-f_1 + \cdots + f_{N-1} - f_{N-3} + f_{N} - f_{N-2} \\
        &= -f_0 - f_1 + f_{N-1} + f_N.
    \end{split}
\end{align}

\subsection{Momentum} \label{subsection:lagrange-multipliers-momentum}

We have $\mathbf{E} = -\mathbf{u} \times \mathbf{B} + \frac{1}{ne} \mathbf{J}\times\mathbf{B} - \frac{1}{ne}\nabla_xp_e$. When we take the first moment of the Vlasov equation, neglecting the spatial derivative, which naturally goes to zero when we integrate over the full domain, we find the time derivative of momentum equal to $\langle \mathbf{v},\mathbf{F}\cdot\nabla_v f\rangle$. We need to be sure this inner product goes to zero, proving the time derivative goes to zero. We have $\mathbf{F} = \frac{q}{m}\left(\mathbf{E} + \mathbf{v}\times \mathbf{B}\right) = \frac{q}{m}\left(\left(-\mathbf{u} \times \mathbf{B} + \frac{1}{ne} \mathbf{J}\times\mathbf{B} - \frac{1}{ne}\nabla_xp_e\right) + \mathbf{v}\times\mathbf{B}\right)$, so the inner product becomes

\begin{equation} \label{eq:inner-product-force-to-three-symmetries}
        m\left\langle \mathbf{v},\mathbf{F}\cdot\nabla_v f\right\rangle = q\left\langle \mathbf{v},\left(-\mathbf{u}\times \mathbf{B} + \mathbf{v}\times\mathbf{B}\right)\cdot\nabla_v f\right\rangle + q\left\langle\mathbf{v}, \frac{1}{ne}\left(\mathbf{J}\times\mathbf{B}\right)\cdot\nabla_v f\right\rangle - q\left\langle\mathbf{v}, \frac{1}{ne}\nabla_xp_e\cdot\nabla_vf\right\rangle.
\end{equation}

The first term on the RHS will be shown to be zero; the last two will be shown to be spatial derivatives scaled by $n$, which go to zero when integrated over the spatial domain. We approach these first. In Section \ref{section:conservative-update-scheme} it was noted that integration by parts results in both of these terms becoming scaled spatial derivatives, which sum to zero when integrated over the domain. However, there is a catch that lies in the fact that $\left\langle\mathbf{v}, \nabla_vf\right\rangle \neq -\left\langle f,\mathbf{1}\right\rangle$ in a finite setting. As such, we find the ratio $\gamma_{v_\eta} \coloneqq -\frac{\left\langle f, 1\right\rangle}{\left\langle v_\eta, \left(\nabla_vf\right)_\eta\right\rangle}$ for $\eta \in \{x,y,z\}$. We see $\left\langle f, 1\right\rangle = -\gamma_{v_\eta}\left\langle v_\eta, \left(\nabla_vf\right)_\eta\right\rangle$, or 
\begin{equation}
    \left\langle \mathbf{1}, f\right\rangle = -\Gamma\left\langle \mathbf{v}, \nabla_vf\right\rangle, \quad \Gamma \coloneqq \text{diag}(\gamma_{v_x}, \gamma_{v_y}, \gamma_{v_z}).
\end{equation}

We can bake this $\Gamma$ term into our force, specifically the electric field $\tilde{\mathbf{E}} \coloneqq \Gamma\mathbf{E}$. So we see \eqref{eq:inner-product-force-to-three-symmetries}, under this electric field, becomes
\begin{align} \label{eq:Gamma-mult-E-terms}
    \begin{split}
        m\left\langle \mathbf{v},\frac{q}{m}\left(\tilde{\mathbf{E}} + \mathbf{v}\times\mathbf{B}\right)\cdot\nabla_v f\right\rangle &= q\left\langle \mathbf{v},\left(-\Gamma\left(\mathbf{u}\times \mathbf{B}\right) + \mathbf{v}\times\mathbf{B}\right)\cdot\nabla_v f\right\rangle \\
        &+ q\left\langle\mathbf{v}, \frac{1}{ne}\Gamma\left(\mathbf{J}\times\mathbf{B}\right)\cdot\nabla_v f\right\rangle - q\left\langle\mathbf{v}, \frac{1}{ne}\Gamma\nabla_xp_e\cdot\nabla_vf\right\rangle.
    \end{split}
\end{align}

By construction the last two terms may be rewritten
\begin{align}
    \begin{split}
        q\left\langle\mathbf{v}, \frac{1}{ne}\Gamma\left(\mathbf{J}\times\mathbf{B}\right)\cdot\nabla_v f\right\rangle - q\left\langle\mathbf{v}, \frac{1}{ne}\Gamma\nabla_xp_e\cdot\nabla_vf\right\rangle &=  -q\left\langle f, \frac{1}{ne}\left(\mathbf{J}\times\mathbf{B}\right) \right\rangle + q\left\langle f, \frac{1}{ne}\nabla_xp_e\right\rangle \\
        &= -\frac{qn}{ne}\nabla_x\cdot\left(\mathbf{B}\mathbf{B} - \frac{1}{2}B^2\right) + \frac{qn}{ne}\nabla_xp_e.
    \end{split}
\end{align}

By quasi-neutrality $\frac{qn}{ne} = 1$, so the coefficient goes away (if the distribution function had multiple species, we would first sum over $s$ to the same result). Upon integrating over space, assuming boundary conditions that cooperate, these terms go to zero, leaving only the first term.

There is a slight catch here. We used a vector calculus identity to acquire $\left(\nabla\times\mathbf{B}\right)\times\mathbf{B} = \nabla\cdot\mathbf{B}\mathbf{B} - \frac{1}{2}\nabla B^2 - \mathbf{B}\left(\nabla\cdot\mathbf{B}\right)$, which is how we converted $\mathbf{J}\times\mathbf{B}$ to a summation of differences which will integrate to zero. While this works in the continuous setting, this does break down in the discrete. So really we have
\begin{equation}
    \mathbf{J}\times\mathbf{B} = \nabla\cdot\mathbf{B}\mathbf{B} - \frac{1}{2}\nabla B^2 + \mathbf{R}_0.
\end{equation}

We can remedy this by simply computing $\mathbf{R}_0$, scaling it by $\frac{1}{ne}$, and subtracting it out from $\mathbf{E}$. In practice, however, this has not been necessary, due to the fact that $\mathbf{J} = \nabla\times\mathbf{B}$ and $\mathbf{B} = \nabla\times\mathbf{A}$, and so the sum over $\mathbf{J}\times\mathbf{B}$ may be rewritten
\begin{equation}
    \sum_i\left(\left(\nabla\times\left(\nabla\times\mathbf{A}\right)\right)\times\left(\nabla\times\mathbf{A}\right)\right)_i = \sum_i\left(\nabla\cdot\left(\left(\nabla\times\mathbf{A}\right)\left(\nabla\times\mathbf{A}\right)\right)\right)_i -\sum_i\left(\left(\nabla\times\mathbf{A}\right)\left(\nabla\cdot\left(\nabla\times\mathbf{A}\right)\right)\right)_i - \frac{1}{2}\sum_i\left(\nabla\lVert\nabla\times\mathbf{A}\rVert^2\right)_i.
\end{equation}

The middle term goes to zero pointwise, with no boundary conditions required, since central differencing preserves $\nabla\cdot\left(\nabla\times\mathbf{a}\right) = 0 \, \forall \, \mathbf{a}$ exactly; the first and third terms are a discrete divergence and a discrete gradient, respectively, so each telescopes and sums to zero assuming appropriate boundary conditions.

When seeking to eliminate the first term in \eqref{eq:Gamma-mult-E-terms} we must use caution. The immediate solution that comes to mind is to compute some Lagrange multiplier analogous to $\gamma_{v_\eta}$, $\lambda_{v_\eta} \coloneqq \frac{\left\langle v_\eta,\left(\mathbf{v}\times\mathbf{B}\right)\cdot\nabla_vf\right\rangle}{\gamma_{v_\eta}\left\langle v_\eta, \left(\mathbf{u}\times\mathbf{B}\right)\cdot\nabla_v f\right\rangle}, \Lambda \coloneqq \text{diag}(\lambda_{v_x}, \lambda_{v_y}, \lambda_{v_z})$, and then apply it to the symmetry, forcing it to zero:
\begin{equation}
    \left\langle\mathbf{v}, \left(-\Lambda\Gamma\left(\mathbf{u}\times\mathbf{B}\right) + \mathbf{v}\times\mathbf{B}\right)\cdot\nabla_vf\right\rangle = 0.
\end{equation}

However, two issues stand in the way. The first, it is common that $\left\langle\mathbf{v},\left(\mathbf{u}\times\mathbf{B}\right)\cdot\nabla_vf\right\rangle$ is extremely small or even zero. The second, even if this divide-by-zero were avoidable, we do not have a handle on the individual components of $\mathbf{E}$, which is passed to us from a black-box method, meaning we would have to apply $\Lambda$ to $\tilde{\mathbf{E}}$ directly. While this would take care of the first term, the last two terms would now be thrown off.

Instead of multiplying by $\Lambda$, let us consider it a pseudo-force that we add to $\tilde{\mathbf{E}}$. We simply compute the difference, rather than the ratio, of the two integrals, normalized by $\langle v_\eta,(\nabla_vf)_\eta\rangle$:

\begin{equation}
    \lambda_{v_\eta} \coloneqq \left(\left\langle v_\eta, -\gamma_{v_\eta}\left(\mathbf{u}\times\mathbf{B}\right)_\eta(\nabla_vf)_\eta\right\rangle + \left\langle v_\eta, \left(\mathbf{v}\times\mathbf{B}\right)_\eta(\nabla_vf)_\eta\right\rangle\right) \big/ \left\langle v_\eta, (\nabla_vf)_\eta\right\rangle, \quad \vec{\lambda} \coloneqq (\lambda_{v_x}, \lambda_{v_y}, \lambda_{v_z})^T.
\end{equation}

We then subtract this from the electric field
\begin{equation}
    \hat{\mathbf{E}} \coloneqq \tilde{\mathbf{E}} - \vec{\lambda}.
\end{equation}

We use this corrected electric field in the update. The pseudo-force will eliminate whatever leftovers exist from the first symmetry.
\begin{align}
    \begin{split}
        \left\langle \mathbf{v},\left(\hat{\mathbf{E}} + \mathbf{v}\times\mathbf{B}\right)\cdot\nabla_v f\right\rangle &= \left\langle \mathbf{v},\left(\left(\Gamma\left(-\mathbf{u}\times \mathbf{B} + \frac{1}{ne} \mathbf{J}\times\mathbf{B} - \frac{1}{ne}\nabla_xp_e\right) - \vec{\lambda}\right) + \mathbf{v}\times\mathbf{B}\right)\cdot\nabla_v f\right\rangle \\
        &= \left\langle \mathbf{v},\left(-\Gamma\left(\mathbf{u}\times \mathbf{B}\right) + \mathbf{v}\times\mathbf{B}\right)\cdot\nabla_v f\right\rangle - \left\langle \mathbf{v},\vec{\lambda}\cdot\nabla_v f\right\rangle
    \end{split}
\end{align}

Componentwise, we see this is really
\begin{align}
    \begin{split}
        &\left\langle v_\eta, \sum_{\alpha\in\{x,y,z\}}\left(-\gamma_{v_\alpha}\left(\mathbf{u}\times\mathbf{B}\right)_\alpha + \left(\mathbf{v}\times\mathbf{B}\right)_\alpha\right)\frac{\partial f}{\partial v_\alpha}\right\rangle - \left\langle v_\eta,\sum_{\alpha\in\{x,y,z\}}\lambda_{v_\alpha}\frac{\partial f}{\partial v_\alpha}\right\rangle \\
        = &\sum_{\alpha\in\{x,y,z\}}\left[\left\langle v_\eta,\left(-\gamma_{v_\alpha}\left(\mathbf{u}\times\mathbf{B}\right)_\alpha + \left(\mathbf{v}\times\mathbf{B}\right)_\alpha\right)\frac{\partial f}{\partial v_\alpha}\right\rangle - \lambda_{v_\alpha}\left\langle v_\eta, \frac{\partial f}{\partial v_\alpha}\right\rangle\right] \\
        = &\sum_{\alpha\in\{x,y,z\}}\left[\left\langle v_\eta,\left(-\gamma_{v_\alpha}\left(\mathbf{u}\times\mathbf{B}\right)_\alpha + \left(\mathbf{v}\times\mathbf{B}\right)_\alpha\right)\frac{\partial f}{\partial v_\alpha}\right\rangle - \frac{\left\langle v_\alpha, -\gamma_{v_\alpha}\left(\mathbf{u}\times\mathbf{B}\right)_\alpha\frac{\partial f}{\partial v_\alpha}\right\rangle + \left\langle v_\alpha, \left(\mathbf{v}\times\mathbf{B}\right)_\alpha\frac{\partial f}{\partial v_\alpha}\right\rangle}{\left\langle v_\alpha, (\nabla_vf)_\alpha\right\rangle}\left\langle v_\eta, \frac{\partial f}{\partial v_\alpha}\right\rangle\right].
    \end{split}
\end{align}

For $\alpha = \eta$, we see the numerator and denominator in the second term cancel, and we are left with

\begin{align}
    \begin{split}
        \left\langle v_{\eta=\alpha},\left(-\gamma_{v_\alpha}\left(\mathbf{u}\times\mathbf{B}\right)_\alpha + \left(\mathbf{v}\times\mathbf{B}\right)_\alpha\right)\frac{\partial f}{\partial v_\alpha}\right\rangle - \left\langle v_\alpha, -\gamma_{v_\alpha}\left(\mathbf{u}\times\mathbf{B}\right)_\alpha\frac{\partial f}{\partial v_\alpha}\right\rangle - \left\langle v_\alpha, \left(\mathbf{v}\times\mathbf{B}\right)_\alpha\frac{\partial f}{\partial v_\alpha}\right\rangle.
    \end{split}
\end{align}

This clearly cancels. However, for $\eta \neq \alpha$, this is more complicated, as the second term disappears entirely after being multiplied by a 0, as $\left\langle v_\beta,\frac{\partial f}{\partial v_\alpha}\right\rangle$ is zero for $\alpha\neq\beta$ assuming zero flux through the boundaries (for similar reasons as mass, we can appeal to the divergence theorem). We are saved, however, by observing the first term of this inner product, $\left\langle v_\eta,-\gamma_{v_\alpha}(\mathbf{u}\times\mathbf{B})\frac{\partial f}{\partial v_\alpha}\right\rangle$ is of the same nature as $\left\langle v_\beta,\frac{\partial f}{\partial v_\alpha}\right\rangle$, given $\mathbf{u}\times\mathbf{B}$ is constant with respect to $\mathbf{v}$. As such, this term goes to zero. Additionally, noting that $(\mathbf{v}\times\mathbf{B})_\alpha$ has no dependence on $v_\alpha$, the term $\left\langle v_\eta, (\mathbf{v}\times\mathbf{B})_\alpha\frac{\partial f}{\partial v_\alpha}\right\rangle$ is again of the same nature as $\left\langle v_\beta, \frac{\partial f}{\partial v_\alpha}\right\rangle$; this, too, goes to zero by appeal to divergence theorem and asserting zero flux boundary conditions.

\subsection{Energy} \label{subsection:lagrange-multipliers-energy}

We have modified $\mathbf{E} \rightarrow \Gamma\mathbf{E} - \vec{\lambda}$, so any modifications for energy will need to come through another source. As such, we will now consider any errors as differences, not as scalings. Four remainder terms will come from our modification of $\mathbf{E}$, with an additional remainder resulting from IBP.

Recall that our total energy update equation is as follows
\begin{equation}
    \frac{\partial}{\partial t}\left[\kappa + (\gamma-1)^{-1}p_e + \frac{1}{2}\frac{1}{\mu_0}B^2\right] = qn\mathbf{u}\cdot\hat{\mathbf{E}} + \mathbf{u}_e\cdot\nabla_xp_e - \mathbf{E}\cdot\mathbf{J}.
\end{equation}

Were $\hat{\mathbf{E}}$ our original electric field, the RHS would go to zero. However, from the momentum correction we have $\hat{\mathbf{E}} = \Gamma\mathbf{E} - \vec{\lambda}$ that now is a part of our kinetic energy, which will cause four remainder terms to appear in the RHS:
\begin{align} \label{eq:R1-R4-remainder-derivation}
    \begin{split}
        &qn\mathbf{u}\cdot\hat{\mathbf{E}} + \mathbf{u}_e\cdot\nabla_xp_e - \mathbf{E}\cdot\mathbf{J} \\
         =&qn\mathbf{u}\cdot\Gamma\left(-\mathbf{u}\times\mathbf{B} - \frac{1}{ne}\nabla_xp_e + \frac{1}{ne}\mathbf{J}\times\mathbf{B}\right) - qn\mathbf{u}\cdot\vec{\lambda} + \mathbf{u}_e\cdot\nabla_xp_e - \left(-\mathbf{u}\times\mathbf{B} - \frac{1}{ne}\nabla_xp_e + \frac{1}{ne}\mathbf{J}\times\mathbf{B}\right)\cdot\mathbf{J} \\
        = &-R_1 - \mathbf{u}\cdot\frac{qn}{ne}\nabla_xp_e - R_2 + \mathbf{u}\cdot\frac{qn}{ne}\mathbf{J}\times\mathbf{B} - R_3 - R_4 + \mathbf{u}_e\cdot\nabla_xp_e + \left(\mathbf{u}\times\mathbf{B}\right)\cdot\mathbf{J} + \frac{1}{ne}\mathbf{J}\cdot\nabla_xp_e \\
        = &\left(-\mathbf{u} + \mathbf{u}_e + \frac{1}{ne}\mathbf{J}\right)\cdot\nabla_xp_e + \mathbf{u}\cdot\left(\mathbf{J}\times\mathbf{B}\right) + \left(\mathbf{u}\times\mathbf{B}\right)\cdot\mathbf{J} - \sum_{\ell=1}^{4}R_\ell \\
        = &-\sum_{\ell=1}^{4}R_\ell
    \end{split}
\end{align}

We define the remainder terms as
\begin{align}
    R_1 &\coloneqq qn\mathbf{u}\cdot\Gamma\left(\mathbf{u}\times\mathbf{B}\right), \\
    R_2 &\coloneqq \frac{qn}{ne}\mathbf{u}\cdot\Gamma\nabla_xp_e - \frac{qn}{ne}\mathbf{u}\cdot\nabla_xp_e = \mathbf{u}\cdot\left(\Gamma-\mathcal{I}\right)\nabla_xp_e, \\
    R_3 &\coloneqq -\frac{qn}{ne}\mathbf{u}\cdot\Gamma\left(\mathbf{J}\times\mathbf{B}\right) + \frac{qn}{ne}\mathbf{u}\cdot\left(\mathbf{J}\times\mathbf{B}\right) = \mathbf{u}\cdot\left(\mathcal{I} - \Gamma\right)\left(\mathbf{J}\times\mathbf{B}\right), \\
    R_4 &\coloneqq qn\mathbf{u}\cdot\vec{\lambda}.
\end{align}

Now we turn to consider the last IBP remainder. In the finite setting, \eqref{eq:IBP-discrete-velocity-appendix} has an error term, i.e.
\begin{equation}
    \left\langle \frac{1}{2}m\left\lVert\mathbf{v}\right\rVert^2,\mathbf{F}\cdot\nabla_vf\right\rangle = -qn\hat{\mathbf{E}}\cdot\mathbf{u} + R_5.
\end{equation}

Taking the contracted energy moment, $\frac{1}{2}m\langle \lVert\mathbf{v}\rVert^2, \cdot\rangle$, of the Vlasov equation yields
\begin{equation}
    \frac{\partial\kappa}{\partial t} + \nabla_x\cdot\mathbf{Q} - qn\mathbf{u}\cdot\hat{\mathbf{E}} + R_5 = 0.
\end{equation}

Now, following the logic of Section \ref{subsubsection:energy-conservation-discrete}, if we sum the time derivatives of the ion kinetic, electron, and magnetic energies together and neglect spatial derivatives, which will vanish upon integration over space, we see
\begin{equation}
    \frac{\partial}{\partial t}\left[\kappa + (\gamma-1)^{-1}p_e + \frac{1}{2}\frac{1}{\mu_0}B^2\right] - qn\mathbf{u}\cdot\hat{\mathbf{E}} + R_5 - \mathbf{u}_e\cdot\nabla_xp_e + \mathbf{E}\cdot\mathbf{J} = 0.
\end{equation}

Following \eqref{eq:R1-R4-remainder-derivation}, we see our full energy equation becomes:
\begin{equation} \label{eq:energy-budget-remainder}
    \frac{\partial}{\partial t}\left[\kappa + (\gamma-1)^{-1}p_e + \frac{1}{2}\frac{1}{\mu_0}B^2\right] = -\sum_{\ell=1}^5R_\ell.
\end{equation}

So we see our spurious source term that comes from discretization errors. We inject $(\gamma-1)\sum_{\ell=1}^5R_\ell$ into the RHS of the pressure update equation, and upon integration over space to eliminate the unwritten spatial derivative terms, we will have restored total energy conservation.

\newpage

\printbibliography

\end{document}